\documentclass[12pt,preprint]{aastex}

\usepackage{apjfonts}
\usepackage{emulateapj5}
\usepackage{amsmath}
\usepackage{psfig}
\usepackage{epsfig}
\usepackage{amssymb}

\renewenvironment{figure}{\begin{figure*} }{\end{figure*}}

\renewcommand{\div}{{\mathbf \nabla} \cdot}

\newcommand{\dd}{{\rm d}}

\begin{document}

\title{Analytical protostellar disk models 1: the effect of 
internal dissipation and surface irradiation on the structure of disks
and the location of the snow line around Sun-like stars}
\author{P. Garaud, D. N. C. Lin} 

\affil{Department of Applied Mathematics and Statistics, Baskin School of Engineering, University of California
Santa Cruz, 1156 High Street, CA 95064 Santa Cruz, USA \\ 
Department of Astronomy and Astrophysics, University of California
Santa Cruz, 1156 High Street, CA 95064 Santa Cruz, USA}

\maketitle

\begin{abstract}
In the classical theory of accretion disks, Lynden-Bell \& Pringle
demonstrated that angular momentum transport and mass diffusion are
associated with the generation of energy through viscous
dissipation. In the seminal model of Chiang \& Goldreich, the only
heating mechanism for the outer regions of protostellar disks is
assumed to be stellar irradiation.  Here, we construct a new set of
self-consistent analytical disk models by taking into account both
sources of thermal energy. We analyze the non-isothermal structure of
the disk across the mid-plane for optically thick disks, and use the
standard two-temperature model in the case of optically thin disks. We
deduce a set of general formula for the relationship between the mass
accretion rate and the surface density profile.  With this
prescription, the evolution of the disk can be studied for a set of
arbitrary initial and boundary conditions if desired. When a Minimum
Solar Nebula surface density profile is prescribed, our results
recover those of Chiang \& Goldreich in the optically thin regions,
but extend their work for the opaque regions of the disk.

For the purpose of illustration, we apply our theory in this paper to
determine the structure of protostellar disks around T Tauri stars
under a state of steady accretion and derive the corresponding radial
variation of disk properties such as surface
density and temperature near the mid-plane. Our model predicts
that the overall photospheric temperature of the opaque disk regions
is typically smaller than previously thought. This modification is 
counter-intuitively related to a reduced flaring angle of the disk 
when viscous heating near the mid-plane is taken into account. 
We calculate the position
of the snow line around a sun-like T Tauri star, and deduce that it
can evolve from well outside 10 AU during FU Orionis outbursts, to $\sim 2$
AU during the quasi-steady accretion phase, 
to the present-day orbital radius
of Venus when the accretion rate falls to about $10^{-9} M_\odot$
yr$^{-1}$, and finally re-expand slightly beyond 2.2 AU during the
protostellar- to-debris disk transition.  This non-monotonous
evolution of the snow line may provide some novel and deterministic
explanation for the origin and isotopic composition
of water on both Venus and the Earth. In addition, in classical T Tauri stars 
 we infer the presence of a marginally opaque region between the optically
thin and thick regions, with a surface density profile varying as
$r^{-3/2}$, as in a Minimum Solar Nebula model. The temperature is 
both vertically and radially constant and up to 40\% higher than that in the
region immediately within.  This transition may lead to an upturn in
the SEDs in the MIR ($24-70 \mu$m) wavelength range. In the optically
thin, outermost regions of the disk we find that the surface density
profile of the dust varies roughly as $r^{-1}$, which is consistent
with mm observations of spatially resolved disk of Mundy et al. (2000)
\end{abstract}


\section{Introduction}

The coplanar orbits of the planets around the Sun inspired Laplace
(1796) to postulate the nebula hypothesis for solar system formation.
The first evidence for these long anticipated protostellar disks was
inferred from continuum IR emission at a level well above the stellar
black-body spectrum for a large fraction of stars in young clusters
(Adams {\it et al.} 1987).  Some flattened structures have also been
resolved with mm imaging (Beckwith \& Sargent 1992).  The IR
and mm excess emissions are attributed to the reprocessing of stellar
light by dust grains.  Since these disks are assumed
to be the cradle of planet formation, the determination of their
structure and evolution has been a central task in the quest to
reconstruct the origin of planetary systems.

Obtaining accurate models of the thermodynamical structure of 
protostellar disks is important both from observational and
theoretical points of view. Firstly, while spatially resolved 
observations are possible for disks around nearby stars, the majority of 
the available data is in the form of spectral energy distributions (SEDs), 
and can only be fully interpreted with a thorough understanding of the 
underlying disk properties. Secondly, the details of the interaction 
between the gas and the condensed heavy elements (in any form from dust 
grains to planets) depend sensitively on the temperature
and pressure gradient of the gas within the nebula. In short, to
understand both observationally {\it and} theoretically the processes
associated with planetary formation requires first and foremost a
better understanding of the basic disk structure. This is the task we
set out to do in this paper.

\subsection{Disk SEDs and grain properties}

In recent years, multi-wavelength observations have provided a rich
data bank of SEDs for most of the
young stellar objects within several nearby clusters (cf Lada {\it et
al.} 2006).  Detailed numerical modeling in terms of reprocessed
stellar irradiation has been successful in reproducing observations 
(Goldreich \& Chiang 1997, Bell 1999, D'Alessio {\it et al.}
2001, Calvet {\it et al.}  2002, Dullemond \& Dominik 2004).  These
analyses show that SEDs are particularly sensitive to three properties
of the disks: 1) the nature of the dust grains (density, size
distribution, physical and chemical composition), 2) the thermal and
chemical structure of the gas, and 3) the global geometrical structure
of the disk such as the inclination, degree of flaring, size,
presence/absence of a hole (Dullemond {\it et al.} 2006).

Current models of dust particles in disks usually assume that they
have the same power-law size distribution as that of the interstellar
grains (Mathis, Rumpl \& Nordsieck 1977, MRN hereafter), 
but with varying upper and
lower cutoffs.  This power law of the MRN distribution implies that
the total cross sectional area is dominated by the contribution from
the smallest grains (below micron size) whereas the large grains ($>1$
mm) contain most of the solid-phase mass of heavy elements. Consequently
the disk SEDs in the wavelength range 
observed by Spitzer are most sensitive to
the small grains, which are tightly coupled to the gas both thermally
and dynamically.

 The objective of this paper is to construct a set of
comprehensive analytical
models of the thermodynamical structure of the disks,
which can easily be utilized for the statistical analysis of large 
SED datasets and a more
physical interpretation of the planet formation environment.

\subsection{Gas-dust interaction}

In parallel with the observational aspects of the study of the disk
SEDs, large theoretical efforts have recently been deployed toward
understanding the growth of small grains as well as the motion of the
larger particles (and small planetesimals) within the primordial solar
nebula (Weidenschilling 1977, Weidenschilling \& Cuzzi 1993). The properties of the
dust grains have recently been thoroughly studied since evidence for
dust growth (D'Alessio {\it et al.} 2001, Shuping {\it et al.} 2003,
McCabe {\it et al.} 2003) or local (spatial) accumulation (Greaves
{\it et al.} 2005) could be the first signs of ongoing planet
formation.  

While
the tiniest dust grains are dynamically very strongly coupled with the
gas and largely follow its evolution, intermediate size objects
(``pebbles'' to ``boulders'') have a different intrinsic motion which
is sensitively dependent on the structure (temperature, pressure
gradient) and dynamical nature (turbulent versus laminar) of the
nebular gas (Morfill \& Voelk 1984, Supulver \& Lin 2000, Youdin \&
Chiang, 2004, Ciesla \& Cuzzi 2006).

It is thought that gas drag causes m-size objects to spiral into the
central star within a few hundred orbital periods 
(Adachi {\it et al.} 1976), but the ubiquitous presence of planets
around at least 10\% of all observed stars (Marcy {\it et al.} 2000)
suggests the existence of a reliable, yet still unidentified mechanism
for the retention of at least a few percent of the total content of
heavy-elements passing through the nebula. Meanwhile, upper limits on
this retention efficiency have independently been set through
the observed homogeneity of the heavy-element abundances 
among sun-like stars in older clusters such as the Pleiades
(Wilden {\it et al.} 2002), the Hyades (Quillen 2002) and IC4665 (Shen
{\it et al.} 2005). These studies suggest that the total
mass of heavy elements retained by the protostellar disks towards 
planet formation is at most a few times that contained in a Minimum 
Solar Nebula  (MSN, hereafter; Hayashi {\it et al.} 1985). 

A self-consistent model of the thermal structure of protostellar disks 
is essential to
understand these rather strongly constrained estimates of the 
heavy-element retention efficiency during planet formation. 

\subsection{Protostellar disk models}

The thermal structure of the gas is determined by a balance between
the generation of heat (through stellar irradiation (Adams {\it et
al.} 1987) and viscous dissipation (Lynden-Bell \& Pringle 1974), its
transport (through radiative diffusion and advective transport) and
finally cooling (through emission by the gas itself, or more
importantly by the dust grains with which the gas is thermally
connected). In existing analytical models of protostellar disks, the
energy source is usually either attributed to viscous dissipation (Lin
\& Papaloizou 1980, 1985, Ruden \& Lin 1986) or stellar irradiation
(Adams {\it et al.} 1987, Chiang \& Goldreich 1997). More recently,
numerical models of the disks have begun to take both sources into
account (Bell 1999, d'Alessio et al. 2001, Dullemond \& Dominik,
2004).

In this paper, we propose a new simple kind of analytical  model to
calculate the two-dimensional $(r,z)$ thermal structure of the disk
taking into account both energy sources. The nature of the model enables 
us to derive analytical formula for all of the disk quantities, and derive 
scaling relations between the disk properties and the host star properties. 

Our work is largely inspired from the seminal paper by Chiang \&
Goldreich (1997) (CG97 thereafter).  As they suggest, most of the
stellar radiation is intercepted by a superheated layer of grains
which re-emits part of it towards the disk mid plane. We also adopt an
isothermal structure in the optically thin regions above the disk
photosphere (in the case where the mid-plane is optically thick) and
in the marginally opaque or optically thin outer regions of the
disk. However, we introduce two major modifications to their work.

Firstly, we model the optically thick regions in more detail by
considering their non-isothermal structure. A warmer mid-plane is a
natural consequence of heating by viscous dissipation 
(in the case of optically thick disks) combined
with  down-gradient transport towards the photosphere. We assume
a polytropic structure for the disk stratification in the direction 
across the mid-plane. This assumption can be shown to be consistent
with some possible heat generation and transport mechanisms, as for
instance in the case of viscously heated convectively unstable
disks. It can also naturally be considered an approximation to the
true stratification of the disk, with a polytropic index to be chosen
from best fits of numerical simulations.

Radiative heating by the central star is well-known to be 
the dominant heat source throughout most disks; nonetheless for completeness 
we also consider regions of the disk in which the dominant heating
process is through viscous heating, which naturally occurs close to
the central star. While this effect is not particularly important for
classical T Tauri stars ($\dot{M} < 10^{-7} M_\odot/$yr) outward of
1AU, the viscously dominated region can extend up to tens of AU for
much larger mass accretion rates.

Secondly, while CG97 specifically assumed an MSN surface density 
profile ($\Sigma_{\rm MSN} = 1000 r_{\rm
AU}^{-3/2}$g/cm$^{2}$) we derive instead self-consistent formula
between the disk structure and the mass accretion rate. The evolution
of the disk can then be studied for a set of arbitrary initial and boundary
conditions if desired. In this paper we restrict 
our study to the case of steady accretion 
(with a radially constant $\dot{M}$) for the sake of simplicity. 
While still being an idealized assumption, this is the
state towards which accretion disks relax should a constant
supply of material be provided near their outer rim. Consequently, it
is a better approximation to the disk structure for the inner regions
of the disk (and presumably out to a few hundred AU) (Lin \&
Papaloizou 1980, 1985) than the standard MSN power law surface density
profile usually used.

\subsection{The importance of the snow line}
The temperature of a typical T Tauri disk decreases from over $1000$K
within a fraction of 1AU to well below $100$K beyond 10AU (see \S6).
In the low pressure and tenuous environment of protostellar disks, ice
grains sublimate between 150K and 170K.  The snow line is generally
defined to be the location where the gas temperature attains these
values (Hayashi {\it et al.} 1985).

As an example of the potential of the new disk model we
propose, we present new results on the evolution of the snow line
around solar-type T Tauri stars. This analysis is much more than
an academic exercise, however, since the snow line is thought to play a
very important role in the formation of planets.

The presence of a snow line somewhere in the disk separates regions
primarily containing refractory silicates from 
those containing volatile ices. Since the relative abundances of
the lighter elements is at least a factor of four larger than that of
metals, the supply of grains as planet building materials is much
richer beyond the snow line. The large difference between the typical
radial velocity of heavy-elements in condensed form and in gaseous
form suggests the possibility of accumulation of material near the
snow line (Takeuchi \& Lin 2005, Ciesla \& Cuzzi 2006). The
thermal ``anomaly'' in the disk structure caused by the presence of
the snow line may also be associated with a flattening of the pressure
gradient, which could slow down the radial migration of the larger
particles, and further enhance the local accumulation of
material. Finally, this boundary may separate the domain of
terrestrial planet versus gas giant planet formation (Ida \& Lin
2004).

Observational evidence for the presence of a snow line in disks around
other stars is scarce. Measurements of the location of the snow line
in the protosolar nebula have been inferred from the actual water
content and suggestions of aqueous alteration in various families of
meteorites (Lunine 2005) to be between 2.5 and 3 AU, within the
asteroid belt.

We will show that while our model naturally reproduces its current location, it
also predicts large excursions in the position and extent of the snow
line during the early stages of the protosolar nebula. We will discuss
the consequences of these findings for the formation and hydration of 
planets in our solar system.

\subsection{Outline of the paper}

The paper is organized as follows. In \S2, we briefly recapitulate the
total energy balance in the disk as a function of the dominant heating
and cooling processes. We largely follow the footsteps of CG97 except
where explicitly mentioned. In particular, we use their prescription
for the absorption and emission properties of the dust grains, which
are both sensitively dependent on the typical wavelength of the
photons considered. This leads to a complex vertical and radial
structure of the disk, which is also presented in \S2.

In \S3 and \S4, we derive analytical formula for the density and
temperature profile in each of the layers for each of the regions
considered, and deduce the two-dimensional structure of the disk. The
methodology followed consists in calculating the
position of the superheated dust grains as a function of 
the total optical depth of
the disk, deducing the flaring angle of the disk and the associated
intercepted stellar flux and equating it with the energy lost through
radiation to deduce the thermal structure of the disk. Once
obtained, we can then evaluate the mass
accretion rate. In the steady-state accretion approximation, the
constant mass accretion rate assumed is then used to calculate
the total surface density of the disk, and to deduce the total optical
depth which closes the system of equations. Naturally, some of these
steps are simplified in the isothermal case, and in the case where the
dominant energy source is viscous heating. By following this method, 
we ensure that the disk structure is determined in a self-consistent manner.

In \S5, we estimate analytically the location of various critical
radii as a function of the accretion rate and the mass of the central
star.  These radii include those separating the radiatively dominated
optically thin, marginally opaque, optically thick and finally viscously
dominated regions of the disk.  We also obtain a rough 
estimate of the radial location
of the snow line from the radius below which the disk mid-plane 
temperature exceeds 160K. 

In \S6, we apply our prescription to compute the structure of disks
with a range of accretion rates around T Tauri stars.  We confirm that
throughout most of the disk, the flaring angle in a steady-state
accretion state increases with radius. In the optically thin region,
we recover the scalings of CG97 for the temperature and scale height
profiles, since these are independent of mass accretion rate. However,
our surface density profile falls off with radius much less steeply 
(as $r^{-1}$)
than that assumed for the MSN. This finding is
consistent with recent observations of spatially resolved disks (Mundy
{\it et al.} 2000, Looney {\it et al. 2003}).  
In the marginally opaque region, we also
coincidentally recover the scalings of CG97 since our model predicts
in this region a surface density profile varying as that
of the MSN. The temperature is interestingly 
found to be significantly larger than in the optically thick region
just within: thus we predict the existence of a ``hot ring'' at radii
$\sim 20-30$ AU around classical T Tauri stars.  In the
optically thick regions our model predicts a significantly flatter disk, 
with notable consequences on the surface density, 
photospheric temperature profile and on the
position of the snow line. We deduce three important results: the snow
line can be very extended, it has a largely non-monotonic evolution,
and can move (for a solar-type T Tauri star) as far in as the current
orbit of Venus.

Finally, we discuss the assumptions of the model in \S7, as well as
its implications for planetary formation and observations of
protostellar disks.

\section{Basic model assumptions and methodology}
\label{sec:1}

\subsection{Basic framework}

Our model follows a standard 1+1D methodology which implicitly assumes
that the disk is thin, with independent vertical (i.e. normal to the
disk plane) and radial structures. The dust, which largely governs the
thermal structure, is initially assumed to be fully mixed with the
gas; this assumption will be relaxed in future work.  In the present
context, we adopt a prescription in which
\begin{equation}
\rho_{\rm d} = Z \rho \mbox{   ,   }
\label{eq:rhod}
\end{equation}
where $\rho$ is the gas density, $\rho_{\rm d}$ is the dust density,
and the ``metallicity'' $Z$ is in principle a slowly varying function of
radius and time. However, note that the magnitude of $Z$ may change
abruptly near transition fronts ({\it eg}, it decreases by a factor of
four as ice grains cross the sublimation line).  We will consider the
details of the disk structure near the snow line in a follow-up
paper. In addition, note that $\rho_{\rm d}$ refers to the heavy-element mass
density contained in small particles only (with settling times which
are much larger than the turbulent diffusion time); there could be a
significant fraction of ``undetectable'' heavy-elements condensed in
larger-size objects, that the model presented here neglects.

The disk has two principal thermal energy sources: radiative heating
(from the central star) and viscous heating (from turbulent 
dissipation associated with momentum transport). In intermediate and
outer regions of typical disks around T Tauri stars, radiative heating
is by far the largest contribution to the total energy budget and
essentially determines the thermal properties of both the grains and
the gas in the disk. In optically thick regions, even when radiative
heating dominates the {\it total} energy budget, viscous heating plays an
important role in maintaining a warmer mid-plane layer, and determines
the vertical stratification of the disk. Finally, much closer to the
central star, viscous dissipation becomes the main agent in heating
the disk.

The effectiveness of the conductive thermal coupling between the gas
and the dust is crucial to the determination of the disk structure. This
coupling depends essentially on the wavelengths considered, as well as
on the collision time-scale between the dust particles and the gas
molecules. CG97 discussed these processes in their seminal work; we
will not repeat them here. 

For computational convenience and reasonable accuracy, 
we assume as they do that the emissivity and opacity of the grains 
interacting a black-body photon spectrum peaked at temperature $T_i$ are
\begin{equation}
\epsilon_i = \left(\frac{T_i}{T_\star}\right)^{\beta} \mbox{ and }
\kappa_i = \kappa_{\rm V} \left(\frac{T_i}{T_\star}\right)^{\beta}
\mbox{ .  }
\end{equation}
When applying the model to T Tauri stars (see \S6), we shall adopt the
value $\beta = 1$ although all the formula are given for arbitrary
values of $\beta$. In the same spirit, we shall also adopt the 
approximate value of $\kappa_{\rm V} = 1 $cm$^2$/g suggested by the work of 
d'Alessio {\it et al.} (2001).

\subsection{Layered structure of the disk}

\begin{figure}[ht]
\epsscale{2}
\plotone{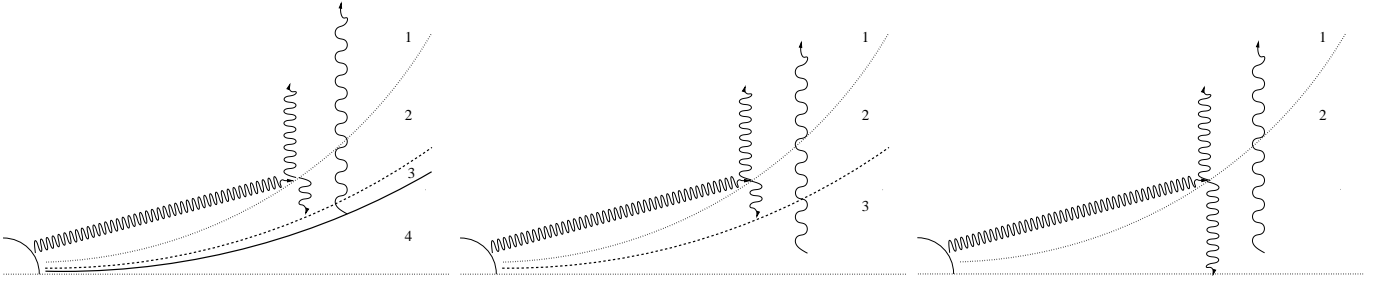}
\caption{Cartoon representation of the layered model for the three regions considered. From left to right, the optically thick, marginally opaque and optically thin cases. In all regions, the top layer
(1) contains superheated grains directly exposed to the (mostly
visible) radiation from the central star. Its lower boundary (the
dotted line) is referred to as $z_{\rm s}(r)$ and is calculated by setting
the optical depth from the central star to the line to unity (see the
Appendix). The superheated grains re-radiate in the IR both towards
infinity and towards the mid-plane. Some of the mid-plane grains are
directly irradiated by the re-processed light from the superheated
grains (layer 2). In the optically thick and marginal cases however, 
part of the mid-plane (layer 3) is optically thick to this emission. 
Grains in layers 2 and/or 3 in turn re-process the absorbed light 
and re-emit it at
slightly longer wavelengths still towards infinity. When the disk is 
optically thick to its own emitted IR
light a fourth layer is constructed, and the emission occurs from the photosphere (solid line) located at $z_{\rm e}(r)$. 
Note that unless the fourth layer is entirely
absent, layer 3 is very thin and can be neglected.}
\label{fig:4zones}
\end{figure}

Following the model of CG97, we identify several layers in a typical
disk.  The detailed structure of these layers in various disk regions
is schematically illustrated in Figure \ref{fig:4zones}. 

The top layer of the disk is referred to as layer 1 where the grains
are directly exposed to the stellar photons. Since sub-mm size dust
grains are inefficient radiators at IR wavelengths, and dust-gas
collisions are too scarce to account for significant cooling, 
grains in layer 1 are ``super-heated''
to a temperature $T_{\rm s}$ satisfying roughly
\begin{equation}
\frac{L_\star}{4\pi r^2} = \epsilon_{\rm s} \sigma T_{\rm s}^4 \mbox{   ,   }
\label{eq:balance1}
\end{equation}
where $L_\star$ is the stellar luminosity, $r$ is the distance from
the central star, $T_{\rm s}$ is the temperature of the superheated
dust grains and $\epsilon_{\rm s}$ is their emissivity at that
temperature.

The energy balance in equation (\ref{eq:balance1}) is valid as long as
the grains are directly heated by the central star. Thus, layer 1
extends from the central star outward until a visual optical depth $1$
is reached (dotted line). The temperature of the gas in this layer is
difficult to determine without a complete description of the
(dynamical, thermal, and chemical) interaction between the dust grains,
the H$_2$ molecules, and the coolant molecules (such as H$_{2}$O, and
others). However, the gas temperature in layer 1 has relatively little
impact on the vertical stratification of the lower regions of the
disk, and for simplicity we shall set it to either the temperature
near the mid-plane $T_{\rm m}$ (in the optically thin case), or the
effective temperature of the disk $T_{\rm e}$ (in the optically thick case).

Super-heated dust grains re-radiate half of the absorbed flux towards 
infinity and half towards the disk mid-plane. 
Where this radiation is absorbed depends on 
the optical depth of the mid-plane layer to re-processed IR photons. 
Thus we construct layers 2 and 3 where layer 2 is optically thin to 
the radiation emitted by the superheated dust-grains, while layer 3 is 
optically thick to it (note that layer 3 only exists for large enough surface 
density). The two layers are separated by the dashed line, defined as the position where the optical depth of the grains to the radiation from the superheated layer reaches unity on a path perpendicular to the midplane.

Finally, the heated mid-plane grains re-emit towards infinity at some
temperature lower than $T_{\rm s}$ still. Again, the total outgoing
flux depends on the optical depth of the mid-plane to its own
emission: if the surface density is large enough, the disk is
optically thick to its own emission and we construct a fourth layer
(layer 4) delimited by the photosphere (solid line).  At the
photosphere, the disk emits a black-body spectrum with temperature
$T_{\rm e}$ which is always lower than $T_{\rm s}$. Owing to the difference
between $T_{\rm e}$ and $T_{\rm s}$ there exist a small difference between the
mid-plane layer photosphere (solid line) and the edge of layers 2 and
3 (dashed line). While this difference is usually negligible (and
neglected) when the inner disk is optically thick, there can exist an
intermediate regime (also discussed by CG97) where the mid-plane layer
is optically thin to its own radiation but optically thick to the
radiation from the superheated layer; in this case, layer 4
vanishes. We refer to this region as the {\it marginally opaque}
region. When the mid-plane layer is also optically thin to the
superheated grains, layer 3 vanishes and we refer to this region as
the {\it optically thin} region. Note that for even smaller total dust content,
 the whole disk becomes optically thin to the radiation from the central star. 
We do not model this last phase in this paper.

\subsection{Variation of the layered structure with radius}
\label{sec:multilayer}

The total energy balance in the various radial regions of the disk is 
now described in more detail.
\bigskip

\noindent {\it Optically-thin disk regions.} \\ 
In the low-surface density regions, mostly at large disk radii, the
entire mid-plane layer is optically thin to the radiation from the
superheated grains as well as to its own emission.  In these regions,
layers 3 and 4 vanish while layer 2 is preserved. The grains in the
mid-plane (now layer 2) are heated by reprocessed radiation from the
superheated grains located on both sides of the disk. As a result, both gas
and individual dust grains have a temperature $T_{\rm m}$ given by 
\begin{equation}
2 \epsilon_{\rm s} \cdot \frac{A_{\rm s}}{2} \frac{L_\star}{4\pi r^2} 
= \epsilon_{\rm m} \sigma T_{\rm m}^4 \mbox{   ,   }
\label{eq:tmthin}
\end{equation}
where $A_{\rm s}$ is the total emitting area-filling factor of superheated
grains above $z_{\rm s}$ ({\it i.e.} the fraction of solid angle covered by
grains), $T_{\rm m}$ is the mid-plane temperature (the gas is assumed to be
isothermal in this region), and $\epsilon_{\rm m}$ is the
corresponding emissivity. Note that the factor of $\epsilon_{\rm s}$
on the LHS accounts for the fact that the grains are inefficient
absorbers at IR wavelengths. We also include the additional factor of
2 accounting for the emission from both layers, which was neglected by
CG97.  
\bigskip
 
\noindent {\it Marginally opaque disk regions.} \\ In an intermediate range of
surface density, the mid-plane layer is optically thick to the
radiation from the superheated dust grains, but optically thin to its
own emission.  In this region, the disk is again assumed to be isothermal
and layer 4 does not exist. Taking into account for the energy budget
for the entire disk, the grain and gas temperature are determined by
\begin{equation}
\frac{A_{\rm s}}{2} \frac{L_\star}{4\pi r^2} = \epsilon_{\rm m} A_{\rm m} \sigma T_{\rm m}^4  \mbox{   ,   }
\label{eq:tmmargin}
\end{equation}
where $A_{\rm m}$ is the total area filling factor of grains in the mid-plane 
layer.
\bigskip

\noindent {\it Opaque disk regions.}  \\
In the limit where the surface density is sufficiently large for the
mid-plane layer to be optically thick to its own radiation, $T_{\rm m}$
cannot be directly determined from the stellar flux. The total energy
budget determines instead the effective temperature $T_{\rm e}$ since the
photosphere is reasonably close to the location where the reprocessed
radiation emitted by the superheated grains is deposited. Thus
\begin{equation}
\frac{A_{\rm s}}{2} \frac{L_\star}{4\pi r^2} = \sigma T_{\rm e}^4  \mbox{   ,   }
\label{eq:tmthick}
\end{equation}
where $T_{\rm e}$ is the temperature (of both dust and gas) at the
photosphere. Note that this equation assumes that the contribution
from viscous heating is negligible, which can be shown to be valid
outward of about 1 AU for a typical T Tauri star.  If we are
interested only in regions outward of the snow-line anyway, this
approximation is justified. However, in the innermost region of the
disk where viscous dissipation is the dominant heating mechanism the
total energy budget becomes
\begin{equation}
\frac{3}{4\pi} \dot{M} \Omega_{\rm K}^2 = 2 \sigma T_{\rm e}^4  \mbox{   ,   }
\label{eq:tmvisc}
\end{equation}
instead, where $\dot{M}$ is the local mass accretion rate, and $\Omega_{\rm K}$
is the local Keplerian angular velocity. This equation relates the
total energy generated by viscous dissipation within the disk
mid-plane to the emitted flux on both sides of the disk.

\subsection{Radial power law: notations}

For simplicity, in each of the four regions, we write the mid-plane 
quantities as 
\begin{eqnarray}
&&  T_{\rm m}(r) = \overline{T} r_{\rm AU}^q  \mbox{   ,   }\\
&& \rho_{\rm m}(r) = \overline{\rho} r_{\rm AU}^p  \mbox{   .   }
\end{eqnarray}
where here and all that follows, we set $r_{\rm AU} = r/(1 \mbox{AU})$.
This uniquely defines the indices $p$ and $q$, in a manner consistent 
with the notation of Takeuchi \& Lin (2002). The resulting isothermal 
disk scale height is therefore obtained as
\begin{equation}
h(r) = \overline{h} r_{\rm AU}^{\frac{q+3}{2}} \mbox{   .   }
\label{eq:hrad}
\end{equation}
Note that $q > -1$ corresponds to a flaring disk, whereas $q < -1$
corresponds to a self-shadowed disk. In all that follows, we assume
that the disk is flaring and check this assumption for
self-consistency.  The disk surface density profile is
\begin{equation}
\Sigma = \overline{\Sigma} r_{\rm AU}^{-s} \mbox{   .   }
\label{eq:sigmarad}
\end{equation}
For instance, one can set $s=3/2$ to recover the MSN model. 
For a steadily accreting disk on the
other hand we determine $\Sigma$ from the requirement that the total
mass accretion rate $\dot{M} = 3\pi \nu_{\rm t} \Sigma$ be constant
with radius, where we have adopted a standard $\alpha$-prescription
for the eddy-viscosity $\nu_{\rm t}$ with value
\begin{equation}
\nu_{\rm t} = \alpha_{\rm t} c_{\rm m} h \mbox{   .   }
\end{equation}
In that case
\begin{eqnarray}
&& s = q + \frac{3}{2} \mbox{   ,   } \\
&& \overline{\Sigma} = \frac{\dot{M}}{3\pi \alpha_{\rm t}} \frac{(1 \mbox{AU})^{3/2}}{\sqrt{\gamma G M_\star} \mbox{   } \overline{h}^2} \mbox{   .   }
\label{eq:steadyacc}
\end{eqnarray}

\section{Structure of optically thin and marginal disk regions}

In the outer regions of the disk, the total surface density of grains
is sufficiently low for the disk to be optically thin to its own
radiation. In this case, layer 4 vanishes. We assume, as in CG97, that
the gas is isothermal in the direction normal to the plane of the disk
and has a temperature $T_{\rm m}$.

Integrating the equation for hydrostatic equilibrium yields the
standard Gaussian vertical density distribution for the gas:
\begin{equation}
\rho(r,z) = \rho_{\rm m}(r) \exp\left(-\frac{z^2}{2h^2(r)}\right) \mbox{   with   } h^2 = \frac{c_{m}^2}{\gamma \Omega_{\rm K}^2}= \frac{R_{\rm g} T_{\rm m}}{\mu \Omega_{\rm K}^2} \mbox{   ,   }
\end{equation} 
where $c_{\rm m}$ is the sound speed for a gas at temperature $T_{\rm m}$, with
mean molecular weight $\mu$ and adiabatic index $\gamma$; $R_{\rm g}$
is the gas constant. When the dust and the gas are fully mixed, the
interface $z_{\rm s}$ between layer 1 and layer 2 (which can be thought of
as the position of the superheated dust layer) is obtained by solving the
equation
\begin{equation}
 \int_0^r \tilde{Z} \rho(r',z') \kappa_{\rm V} \dd {\bf l} = 1 
\end{equation}
on a straight path $\dd {\bf l}$ from the central star to the point
$(r,z_{\rm s})$ (namely, $z' = (z_{\rm s}/r) r'$). Here, $\tilde{Z}$ is the
assumed metallicity $Z$ normalized to the value used for the
determination of $\kappa_{\rm V}$, namely 1\%. Thus $\tilde{Z} =
Z/0.01$. Note that it can also be regarded as $\sim 10^{\rm [Fe/H]}$
where [Fe/H] is the metallicity normalized with respect to that of the
Sun. Assuming the disk is flaring, one can approximate this equation
using the variant of a stationary phase method (see Appendix), which
yields the equation (valid at every radius $r$ within the region)
\begin{equation}
\tilde{Z} \rho_{\rm m} h \kappa_{\rm V} \exp\left(-\frac{z_{\rm s}^2}{2h^2}\right) = \frac{z^2_{\rm s}}{h^2} \alpha + \frac{h}{r} \frac{\dd \ln (\tilde{Z}  \rho_{\rm m})}{\dd \ln r} \mbox{   ,   }
\label{eq:tau0}
\end{equation}
where we have defined $\alpha$ as the grazing angle between the incoming 
radiation from the star and the local stratification:
\begin{equation}
\alpha = r\frac{\dd}{\dd r} \left(\frac{h}{r} \right) \mbox{   .   }
\label{eq:alpha}
\end{equation}

Note that this approximation is only strictly valid in the limit where
$z_{\rm s}$ is larger than a few scale heights. By inspection, one can note
that the second term on the RHS of equation (\ref{eq:tau0}) is
negligible within the same limit unless the disk has a sudden jump in gas density or metallicity (this could occur near a planet-induced gap, or near the successive sublimation zones). Deferring the study of these transition zones to a later paper, we find that the ratio $z_{\rm s}/h$ is otherwise given
by
\begin{equation}
\frac{z_{\rm s}}{h} = \left[ 2 {\rm LambertW}
\left(\frac{\tau^{\rm tot}_{\rm V} }{\sqrt{2\pi} \alpha} \right) 
\right]^{1/2} \mbox{   ,   }
\end{equation}
where $\tau_{\rm V}^{\rm tot} = \tilde{Z} \Sigma \kappa_{\rm V}/2$ is
the total visual optical depth from the mid-plane to infinity on a
path perpendicular to the disk. Given that the disk is isothermal,
$\Sigma = \sqrt{2\pi} \rho_{\rm m} h$.

\begin{figure}
\epsscale{1}
\plotone{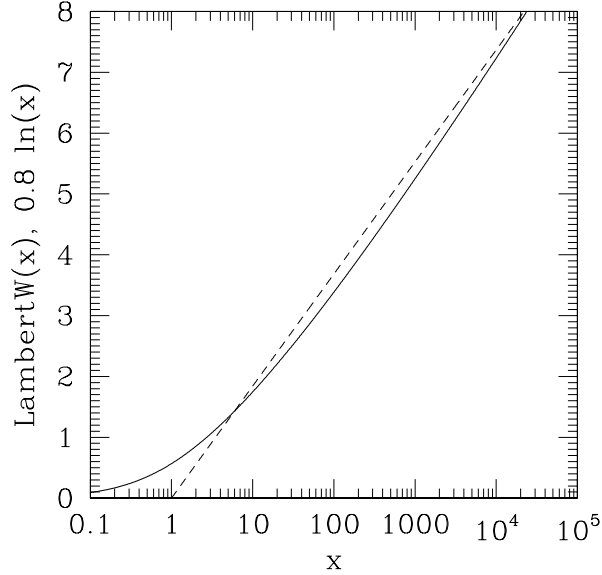}
\caption{LambertW(x) function (solid line) and its approximation
$0.8 \ln(x)$ (dashed line) for large values of $x$. Although 
this is not an asymptotic approximation, and is not uniformly 
valid for $x \rightarrow \infty$, it is quite good in the range considered.}
\label{fig:LW}
\end{figure}

Note that the LambertW function is defined as the inverse of the
function $f(x) = x\exp(x)$, or in other words, if $y = x\exp(x)$ then $x =
{\rm LambertW}(y)$. For ease of computation, and for large values of
$y$ (for $y$ between 10 and $10^5$ say), a fairly good approximation
(to within about 10\%) of the LambertW function is given by
\begin{equation}
{\rm LambertW}(y) \simeq 0.8 \ln (y) \mbox{   .   }
\end{equation}
This can be seen in Figure \ref{fig:LW}. 

We therefore note that as long as $\tau^{\rm tot}_{\rm V} \gg
\sqrt{2\pi}\alpha$ the ratio $z_{\rm s}/h$ is a slowly varying function of
radius with
\begin{equation}
\frac{z_{\rm s}}{h} = C(r) \simeq \left[ 3.7 \log_{10} 
\left(\frac{\tau^{\rm tot}_{\rm V} }{\sqrt{2\pi} 
\alpha} \right) \right]^{1/2} \mbox{   .   }
\label{eq:C(r)}
\end{equation}
Calculating $C(r)$ exactly in the case of a classical T Tauri star reveals 
that $C$ varies between 3 and 4 in the outer regions
of the disk, confirming the ansatz ($z_{\rm s} \simeq 4 h$) of
CG97. However, for other stellar types or for other mass accretion
rates, we find that the superheated dust layer position varies widely
(with $C$ taking any value between 1 and 6).

Once the magnitude of $z_{\rm s}$ is determined, we need to identify
the surface area filling factor $A_{\rm s}$ of the reprocessing
material contained within the superheated layer which is equal to the
visual optical depth $\tau_{\rm V}^s$. This is done through the
integral
\begin{equation} 
A_{\rm s} \simeq \int_{z_{\rm s}}^\infty \tilde{Z} \rho \kappa_{\rm V} \dd z 
= \tau^s_{\rm V}  \mbox{   .   }
\end{equation}
Using the same stationary phase approximation (see Appendix), we 
evaluate the integral to get
\begin{equation}
A_{\rm s} \simeq \tilde{Z} \rho_{\rm m} \kappa_{\rm V} h \exp\left(-\frac{z_{\rm s}^2}{2h^2}\right) \frac{h}{z_{\rm s}}  \mbox{   .   }
\label{eq:tauvert}
\end{equation}
and combining this with equation (\ref{eq:tau0}) yields
\begin{equation}
A_{\rm s} \simeq \frac{z_{\rm s}}{h} \alpha \mbox{   .   }
\label{eq:zsh}
\end{equation}
Note that this expression differs from the one suggested by CG97 by a
geometrical factor of $z_{\rm s}/h = C(r) $; this difference arises in
choosing to determine $A_{\rm s}$ from a path that is perpendicular to the
mid-plane rather than perpendicular to the disk stratification. We
prefer the former interpretation as more consistent with the 1+1D
assumption, although the latter is also acceptable within the degree
of approximation associated with this analytical approach.

\subsection{Optically thin disk regions}

To compute the radial structure of the disk, we combine equations
(\ref{eq:tmthin}), (\ref{eq:zsh}), and (\ref{eq:alpha}), and ignore
the slow dependence in $r$ of $C(r)$. This derivation yields an
equation for $h(r)/r$ for instance, of the general form described and
solved in the appendix:
\begin{equation}
\frac{\dd}{\dd r} \left(\frac{h}{r}\right) = \frac{1}{C} \left(\frac{\mu}{R_g} 
\frac{GM_\star}{T_\star}\right)^{4+\beta} R_\star^{-\frac{4\beta+8}{4+\beta}} 
h^{8+2\beta} r^{\frac{3\beta+4}{4+\beta}-3(4+\beta)} \mbox{   ,   }
\end{equation}
which can be integrated to yield equation (\ref{eq:hrad}) with 
\begin{eqnarray}
\frac{\overline{h}}{\mbox{AU}} &=& \left[ \frac{1}{C}\frac{2\beta^2 + 15\beta + 28}{\beta^2 
+ 4\beta + 8} \left(\frac{\mu}{R_g} \frac{GM_\star}{R_\star T_\star}\right)^{4+\beta} \right]^{-\frac{1}{7 + 2\beta}} \nonumber \\ \cdot && \left(\frac{R_\star}{1\mbox{AU}}\right)^{-\frac{8+4\beta+\beta^2}{28+15\beta+2\beta^2}}  \nonumber
\end{eqnarray}
\begin{equation}
\frac{q+3}{2} = \frac{\beta^2 + 4\beta + 8}{2\beta^2 + 15\beta  +28} + 1\mbox{   .   }
\label{eq:qthin}
\end{equation}
If $\beta = 1$ then $q = -19/45$, with $h \propto r_{\rm AU}^{58/45}$ and
the flaring angle $\alpha \propto r_{\rm AU}^{13/45}$. This result is
independent of the surface density distribution of the gas, and is
similar (within factors of order unity) to the expression derived by
CG97. If we require in addition that the disk be in a state of steady
accretion, the surface density distribution can be deduced directly
from (\ref{eq:sigmarad}), and in that case $s = 97/90 \simeq 1$. This
result is particularly interesting in the light of the observational
results of Mundy {\it et al.} (2000), who found that the surface
density profile of spatially resolved disks varied as $r_{\rm AU}^{-1}$ rather
than $r_{\rm AU}^{-3/2}$.

This analysis is only valid in the region delimited from the inside by
the inequality $\tau^{\rm tot}_{\rm s} < 2/3$ (where $\tau^{\rm
tot}_{\rm s} = \tilde{Z} \Sigma \kappa_{\rm s}/2$), and from the
outside by the inequality $\tau^{\rm tot}_{\rm V} \gg \alpha$ (or $h
\simeq r$, whichever happens closest to the central star). Note that
as the total optical depth drops below $\alpha$, the superheated layer
becomes progressively more and more optically thin to the radiation
from the central star. This is expressed mathematically by $C
\rightarrow 0$. When this happens, all the grains directly exposed
become ``superheated''. The temperature of the gas, however, is more
difficult to evaluate because of the long collision time scale between
grains and gas molecules; as a result, it is unclear whether the disk
retains its flaring structure or becomes self-shadowed, although we
suspect the latter is more likely. This transition 
would appear as an outer edge to the observable disk.

In addition, when considering the case of steady-state accretion, it
is important to bear in mind the underlying assumptions that (1) the
disk has time to relax to the quasi-steady state, and (2) that a
constant mass supply is indeed provided to the disk from the outer
rim. Both these assumptions are difficult to justify beyond a few
hundred AU.

\subsection{Marginally opaque disk regions}

To compute the radial structure in the marginally opaque case, we need to 
determine the emitting surface of grains in the mid-plane region
$z \in [0,z_{\rm s}]$ which is very well approximated by $\tau_{\rm V}^{\rm
tot}$ in the limit where $z_{\rm s}$ is larger than a few scale heights
(which we proved to be self-consistently true):
\begin{equation}
A_{\rm m} \simeq \tau_{\rm V}^{\rm tot} = \frac{\tilde{Z} \Sigma \kappa_{\rm V}}{2} 
\mbox{   .   }
\label{eq:am}
\end{equation}  
Using the same method as in the previous section (which involves using 
equations (\ref{eq:tmmargin}), (\ref{eq:zsh}) and (\ref{eq:am})), we 
find that the disk scale height is 
\begin{eqnarray}
&& \frac{\overline{h}}{1\mbox{AU}} = \left[ \frac{2}{C} \frac{7 + 2\beta}{2 + s + \beta} 
\frac{\tilde{Z} \overline{\Sigma} \kappa_{\rm V}}{2} \left(\frac{\mu}{R_g} 
\frac{GM_\star}{R_\star T_\star}\right)^{4+\beta} \right]^{-\frac{1}{7 
+ 2\beta}} \left(\frac{R_\star}{1 \mbox{AU}}\right)^{-\frac{2+\beta}{7+2\beta}} \nonumber \\
&& \frac{q+3}{2} = \frac{9+s+3\beta}{7+2\beta}\mbox{   .   }
\label{eq:margin}
\end{eqnarray}

Interestingly, we find that the MSN scaling law is the one 
consistent with steady accretion, with $h \propto r_{\rm AU}^{3/2}$ and
\begin{equation}
s = \frac{3}{2} \mbox{   ,   } q = 0\mbox{   ,   } \alpha = \frac{h}{2r} 
\mbox{   ,   }
\label{eq:alphathin}
\end{equation}
for any value of $\beta$. This value of $q$ implies that the disk
is radially isothermal in this region. In addition,
$\overline{\Sigma}$ and $\overline{h}$ are obtained by solving
simultaneously equations (\ref{eq:margin}) and (\ref{eq:steadyacc}),
so that
\begin{equation}
\frac{\overline{h}}{1\mbox{ AU}} = \left[ \frac{2}{C} \frac{\dot{M}\tilde{Z} \kappa_{\rm V}}
{3\pi\alpha_{\rm t} \sqrt{\gamma G M_\star R_\star}}  \left(\frac{\mu}{R_g} 
\frac{GM_\star}{R_\star T_\star}\right)^{4+\beta} 
\right]^{-\frac{1}{5 + 2\beta}} \left(\frac{R_\star}{1 \mbox{AU}}\right)^{-\frac{1}{2}} \mbox{   ,   }
\end{equation}
and the value of the radially constant temperature is 
\begin{equation}
T_{\rm m} = \frac{\mu}{R_g} \frac{GM_\star}{(1 \mbox{AU})^{3}} \overline{h}^2   \mbox{   .   }
\end{equation} 
Again, this result recovers the expression obtained by CG97. 
The marginally opaque region terminates at the
radius where $\tau^{\rm tot}_{\rm m} = 2/3$, where $\tau^{\rm
tot}_{\rm m} = \tilde{Z} \kappa_{\rm m} \Sigma/2$.

\section{Vertical structure of opaque disk regions}

Disks are heated both by viscous dissipation and by
stellar irradiation. When the mid-plane is opaque
to its own thermal radiation, heat generated near the mid-plane can
only be transported towards the photosphere by turbulent or
radiative down-gradient diffusion. This implies the emergence of a
temperature gradient across the disk mid-plane, which must be
calculated self-consistently. 

\subsection{Thermal stratification in the optically thick layer}

Existing detailed disk models are constructed based on the numerical
integration of this thermal equilibrium equation (cf d'Alessio {\it et
al.}  2001; Dullemond \& Dominik, 2004)
\begin{equation}
\rho \nu \left( r \frac{\partial}{\partial r} (\Omega_{\rm K})
\right)^2 = \div (F_{\rm rad} + F_{\rm turb}) \mbox{   .  }
\end{equation}
Unfortunately the nonlinearities in the radiative contribution to the heat flux
prevent simple analytical solutions. Instead, we adopt a polytropic 
structure for the disk which can either be deduced exactly assuming idealized 
heat transport mechanisms (such as effective convective transport leading 
to an adiabatic stratification), or considered an approximation to the 
solution of the full transport equation.

Let's consider the standard polytropic equation 
\begin{equation}
p = K \rho^{1+1/n}
\end{equation}
where $p$ is the gas pressure, $n$ is the polytropic index and $K$ is the polytropic constant. The equation of hydrostatic equilibrium across the mid-plane is
\begin{equation}
\frac{{\rm d} p}{{\rm d}z} = - \rho \Omega_{\rm K}^2 z
\end{equation} 
where we have neglected the self-gravity of the disk. Combining these with a perfect gas equation of state 
\begin{equation}
p = \frac{R_g \rho T}{\mu}
\end{equation}
where $R_g$ is the gas constant and $\mu$ is the mean molecular weight, we obtain
\begin{eqnarray}
&& T(r,z) = T_{\rm m}(r) \left(1-\frac{z^2}{H(r)^2}\right) \mbox{   ,  } \nonumber \\
&& \rho(r,z) = \rho_{\rm m}(r)  \left(1-\frac{z^2}{H(r)^2}\right)^n \mbox{   ,  }
\end{eqnarray}
where $H$ is uniquely related to the conventionally defined disk isothermal 
scale height
\begin{equation}
h^2 = \frac{c_{\rm m}^2}{\gamma\Omega_{\rm K}^2} \mbox{   ,  }
\label{eq:h00}
\end{equation}
(where $\gamma$ is the adiabatic index for the gas) as
\begin{equation}
\frac{H}{h} = \theta = \sqrt{2n+2} \mbox{   ,} 
\label{eq:z00}
\end{equation}
Note that if the gas is composed of mostly molecular hydrogen with a solar composition,
\begin{equation}
\gamma = 1.4  \mbox{   ,   }  \mu =  2.4 \mbox{   . }
\end{equation} 
If we consider a convective polytrope for instance, we choose $n$ such that $\gamma = 1 + 1/n$ or in other words, $n=2.5$. 

Note also that $H$ determines the location where, in principle, both
the gas density and temperature would drop to zero. However, this
mid-plane solution is only valid within the optically thick interior
of the disk, i.e within the interval $z \in [0,z_{\rm e}(r)]$, where
$z_{\rm e}(r)$ is the position of the photosphere.  For the convenience of
following discussions, we define the ratio
\begin{equation}
\Delta_{\rm e} =z_{\rm e}/H \mbox{   .  }
\end{equation}
Above the photosphere, the temperature and density profiles are
matched onto an isothermal atmosphere solution with the photospheric 
temperature $T_{\rm e}$. The matching procedure involves solving 
the total energy 
equation to determine $T_{\rm e}$, which then uniquely
yield the values of the mid-plane temperature and density $T_{\rm m}$ and
$\rho_{\rm m}$. In order to do this, we must first determine the position of
the photosphere, which can be done once the vertical stratification in
the atmosphere is calculated.

\subsection{Disk atmosphere}

Above the photosphere, we assume the gas to be isothermal with
temperature $T_{\rm e}$, and constant sound speed matched to the 
interface value:
\begin{equation}
c^2(r,z) \equiv c^2_{\rm e}(r) = c^2(r,z_{\rm e}) =
\frac{\gamma R_{\rm g} T_{\rm e}(r)}{\mu} \mbox{   ,  }
\end{equation}
for $z > z_{\rm e}$. Integrating the equation for hydrostatic
equilibrium we obtain the gas density:
\begin{equation}
\rho = \rho_{\rm m}\left(1-\Delta_{\rm e}^2\right)^n  
\exp\left(-\frac{z^2-\Delta_{\rm e}^2 H^2}{2h_{\rm e}^2}\right) \mbox{   ,  }
\end{equation}
where 
\begin{equation}
h^2_{\rm e} = \frac{c_{\rm e}^2}{\gamma \Omega_{\rm K}^2} =
\frac{c_{\rm m}^2}{\gamma \Omega_{\rm K}^2} (1-\Delta_{\rm e}^2) =
h^2 (1-\Delta_{\rm e}^2) \mbox{   .  }
\label{eq:he}
\end{equation}

\subsection{Location of the photosphere.}

To obtain $z_{\rm e}$, we must solve the equation
\begin{equation}
\int_{z_{\rm e}}^{\infty} \tilde{Z} \rho \kappa_{\rm e} \dd z = \frac{2}{3}\mbox{   ,  }
\end{equation}
where $\kappa_{\rm e} = \kappa_{\rm V} (T_{\rm e}/T_\star)^\beta$.
Assuming that the dust is fully mixed with the gas (see equation
(\ref{eq:rhod})) yields the required equation for $\Delta_{\rm e}$:
\begin{eqnarray}
&& \left( 1 -
\Delta_{\rm e}^2\right)^{n+\beta+1/2} {\rm erfc}\left(\frac{\theta
\Delta_{\rm e}}{\sqrt{2-2\Delta_{\rm e}^2}}\right)
\exp\left(\frac{\Delta_{\rm e}^2\theta^2}{2-2\Delta_{\rm e}^2}\right) \nonumber \\ &&
= \frac{2}{3} \sqrt{\frac{2}{\pi}} \frac{1}{\tilde{Z} \rho_{\rm m} h \kappa_{\rm V}} \left(\frac{T_{\rm m}}{T_\star}\right)^{-\beta}\mbox{   .  }
\label{eq:nasty}
\end{eqnarray}
This expression for $\Delta_{\rm e}$ in terms of $\rho_{\rm m}$ and $T_{\rm m}$ at a given
radius $r$ seems rather complicated and requires in principle a
numerical solution. However, there exist two limits in which a simple 
analytical solution exists. In the strongly opaque (when the 
right-hand-side of equation (\ref{eq:nasty}) is very small) we
expect $z_{\rm e}$ to be located high above the mid-plane and very
close to $H$ (so that $\Delta_{\rm e} \simeq 1$). The nature of the 
polytropic/isothermal solution we
are using implies that in this case, we expect
\begin{equation}
1-\Delta_{\rm e}^2 = \varepsilon_{\rm e} \ll 1 \mbox{ so that } \Delta_{\rm e} \simeq 
1-\frac{\varepsilon_{\rm e}}{2} \mbox{   .  }
\label{eq:deltaa}
\end{equation}
Substituting this ansatz into equation (\ref{eq:nasty}), and using the
approximation (\ref{eq:erfc}) we obtain
\begin{equation}
\varepsilon_{\rm e} \simeq \left( \frac{(2/3)\theta^2 I(n)
}{\tau^{\rm tot}_{\rm V}} \right)^{\frac{1}{n+\beta+1}} \left(\frac{T_{\rm m}}{T_\star}\right)^{-\frac{\beta}{n + \beta + 1}} \mbox{   ,  }
\label{eq:epsilon}
\end{equation} 
since in the limit where $z_{\rm e} \simeq H$ the total surface density of the gas can be approximated by 
\begin{equation}
\Sigma(r) = \int_{-\infty}^\infty \rho(r,z) \dd z \simeq
\int_{-H}^{H} \rho(r,z)\dd z = 2 \rho_{\rm m}(r) H(r) I(n) \mbox{   ,  }
\end{equation}
where 
\begin{equation}
I(n) = \int_0^1 (1-x^2)^n \dd x = \frac{1}{2}
B\left(\frac{1}{2},n+1\right) \mbox{   ,  }
\label{eq:idd}
\end{equation}
($B$ is a Beta function, see Gradshteyn \& Rhyshik p. 343). From equation 
(\ref{eq:epsilon}), we note that the approximation 
$\tau^{\rm tot}_{\rm V} \gg 1$ is consistent with $\varepsilon_{\rm e} \ll 1$. 

In the alternative limit, the right-hand-side of equation
(\ref{eq:nasty}) approaches 1 (the disk is close to being optically thin), and
we expect $\Delta_{\rm e}$ to be very small, in which case an approximate
solution can be obtained by linearizing (\ref{eq:nasty})
\begin{equation}
\Delta_{\rm e} \simeq \sqrt{\frac{\pi}{2}} \frac{1}{\theta} 
\left[ 1-\frac{2/3}{\tau^{\rm tot}_{\rm V}} 
\left(\frac{T_{\rm m}}{T_\star}\right)^{-\beta} \right]  \mbox{   ,  }
\label{eq:nearlythin}
\end{equation}
and using $\Sigma \simeq \sqrt{2\pi} \rho_{\rm m} h$. 

\begin{figure} 
\epsscale{1}
\plotone{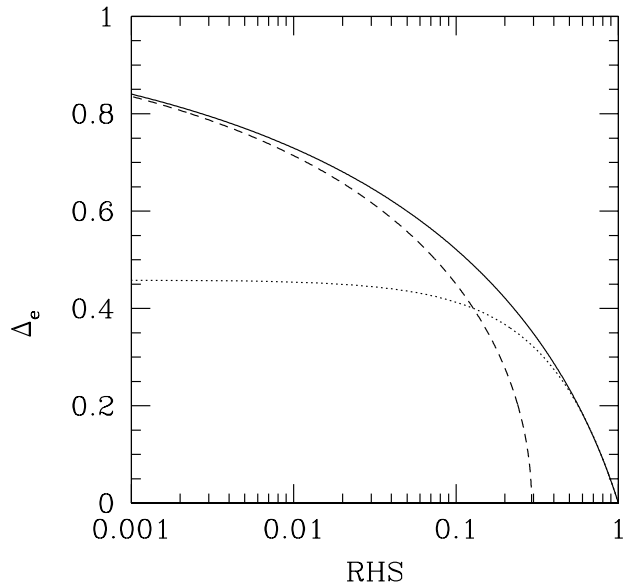}
\caption{Comparison between the true solution to equation
(\ref{eq:nasty}) and the two asymptotic approximations. The x-axis
represents the right-hand-side of equation (\ref{eq:nasty}). The solid
line is the numerical solution, the dotted line is the approximation
in the weakly opaque limit, while the dashed line is the
approximation in the strongly opaque limit.}
\label{fig:nasty}
\end{figure}

The regimes of validity of the two approximations are illustrated in
Figure \ref{fig:nasty}. For the purpose of computations, we will
denote from now on the ``strongly opaque regime'' as the region where
we adopt the asymptotic approximation (\ref{eq:epsilon}), and the
``weakly opaque regime'' as the region where we adopt
(\ref{eq:nearlythin}) instead. Since we are not performing a rigorous
matching between the two regions, we artificially patch them instead
at the radius where the RHS of (\ref{eq:nasty}) reaches 0.1.

\subsection{Characteristics of the superheated dust layer}

The calculation of the position of the superheated layer is similar to
the isothermal case, but with a different vertical density
distribution for the gas. As before, we use the stationary phase
approximation assuming that the disk is flaring.  In the weakly opaque
limit, we retain the expressions derived in the optically thin case
(namely equations (\ref{eq:C(r)}) and (\ref{eq:zsh})), whereas in the
strongly opaque limit (see Appendix), we get
\begin{equation}
z_{\rm s} \simeq H  = \theta h \mbox{   ,  }
\label{eq:thickzs}
\end{equation}
and as before
\begin{equation}
A_{\rm s} = \frac{z_{\rm s}}{h} \alpha \simeq \theta \alpha \mbox{   .  }
\label{eq:thickAs}
\end{equation}
It is interesting to note at this point that in a completely
isothermal model (as the one proposed by CG97), $C(r)$ increases
monotonically with $\tau_{\rm V}^{\rm tot}$ (equation (\ref{eq:C(r)})). 
Here instead we show that $C(r)$ enters a different
regime in the optically thick region, and has a constant value set
uniquely by the gas properties contained in $\theta$ (see equation 
(\ref{eq:z00})).

\subsection{Radial structure}

{\it Weakly opaque region.} Now that we have all the required
information, we need to solve the total energy conservation
equation. When the disk is nearly optically thin to its own
radiation, the mid-plane temperature is very close to the photospheric
temperature: the disk is nearly isothermal (typically, $\varepsilon_{\rm e}
\simeq 1$). In this case, we approximate, for simplicity, the total
energy equation by
\begin{equation}
\frac{C}{2} r \frac{\dd }{\dd r} \left(\frac{h}{r}\right)
\frac{R_\star^2}{ r^2} = \left(\frac{T_{\rm e}}{T_\star}\right)^4 \simeq
\left(\frac{T_{\rm m}}{T_\star}\right)^4 \mbox{   .   }
\end{equation}
Integrating this equation for the disk scale height yields
\begin{eqnarray}
&& \frac{\overline{h}}{1\mbox{AU}} = \left( \frac{C}{7} \right)^{1/7} \left(
\frac{\mu}{R_{\rm g}} \frac{G M_\star}{R_\star T_\star}\right)^{-4/7} \left(\frac{R_\star}{1 \mbox{AU}}\right)^{-2/7}\\ &&
\frac{q+3}{2} = \frac{9}{7} \mbox{   ,   }
\end{eqnarray}
so that $q = -3/7$ and for a steadily accreting disk, $s=15/14$. This
result, once more, naturally recovers the radial power law scaling
found by CG97 since their work assumes that even optically thick disks
are isothermal. Interestingly, we find that $\alpha = 2/7 (h/r)$ which
is notably smaller than the flaring angle in the nearby marginally
opaque region (see equation (\ref{eq:alphathin}).  This difference
indicates that the inclusion of the viscous dissipation reduces the
flaring angle compared to a purely radiatively heated disk. 
This result is not surprising. The scaleheight of a standard disk 
is determined by the balance between the pinching force of gravity, and the vertical pressure gradient which is related to the local temperature. Thus the flaring angle is related to the radial variation of the ratio of these two quantities. In the case of a radially isothermal disk, the flaring angle is maximal since gravity quickly decreases outward. In the case of radiatively heated disks, the addition of a negative temperature gradient implies that temperature increases inward, and reduces the flaring. Finally, the additional contribution from viscous heating to the mid-plane temperature further reduces the flaring. 

What is more counter-intuitive is that the inclusion of viscous dissipation can, in some regions, reduce the photospheric temperature of a disk compared to a model in which it is not taken into account: {\it additional heating does not necessarily mean apparently hotter disks}. This happens in regions where viscous heating is only important in maintaining a warmer mid-plane but does not contribute much to the overall energy budget. In this case, the reduced flaring angle implies a reduced intercepted stellar flux, and therefore a cooler photospheric temperature. 

{\it Strongly opaque regions} 
In what follows, we show how our formalism is able to extend the model
of CG97 to non-isothermal, strongly opaque disks. In this new limit,
we use the fact that $T_{\rm e} = \varepsilon_{\rm e} T_{\rm m}$ with equation
(\ref{eq:epsilon}). After a few manipulations we get
\begin{eqnarray}
\frac{ \overline{h}}{1\mbox{AU}} &=& \left[ \frac{2n+2-4s-2\beta}{7n+7-\beta}
\frac{\theta}{2} \right]^{\frac{n+\beta+1}{7n-\beta+7}}
\left( \frac{\mu}{R_{\rm g}} \frac{G M_\star}{R_\star T_\star}\right)^{-
\frac{4n+4}{7n+7-\beta}} \nonumber \\ &\cdot& \left( \frac{4}{3}
\frac{\theta^2 I(n)}{\tilde{Z} \overline{\Sigma} \kappa_{\rm V}}
\right)^{-\frac{4}{7n+7-\beta}} \left(\frac{R_\star}{1 \mbox{AU}}\right)^{-\frac{2n+2-2\beta}{7n+7-\beta}} \mbox{   ,   } \\ \frac{q+3}{2} &=&
\frac{9n+9-3\beta-4s}{7n+7-\beta} \mbox{   .   }
\end{eqnarray}

In the case of a steadily accreting disk, we combine these equations
with equation (\ref{eq:steadyacc}) to obtain
\begin{eqnarray}
\frac{\overline{h}}{1 \mbox{ AU}} &=& \left[ \frac{2n-2\beta}{7n+15-\beta} \frac{\theta}{2} \right]^{\frac{n+\beta+1}{7n-\beta+15}} \left(
\frac{\mu}{R_{\rm g}} \frac{G M_\star}{R_\star T_\star}\right)^{-
\frac{4n+4}{7n+15-\beta}} \nonumber \\  &\cdot & \left( \frac{4 \pi
\alpha_{\rm t} \sqrt{\gamma G M_\star R_\star} \theta^2 I(n) }{\tilde{Z}  \kappa_{\rm V} \dot{M}} \right)^{-\frac{4}{7n+15-\beta}} \left(\frac{R_\star}{1 \mbox{AU}}\right)^{-\frac{2n-2\beta}{7n-\beta+15}} \mbox{   ,   }  \\  s &=& \frac{15n + 15
- 9\beta}{2(7n + 15 - \beta)} \mbox{   .   }
\end{eqnarray}
With $\beta = 1$, and $n = 2$, we obtain $s \simeq 0.65$ and $q
\simeq -0.85$, which confirms again that the disk is flaring. However,
we note that in this case the flaring is even weaker ($\alpha
\simeq 0.1 h/r$), and the disk surface is very nearly flat. Again, this
geometry has very important consequences for the overall inner disk
temperature, since a much lower fraction of the stellar luminosity is
intercepted than in a model in which the disk is assumed to be
isothermal (see the previous case). 

{\it Viscously heated disks} 
Finally, if the dominant contribution to the heating of the disk arises 
from viscous dissipation instead of stellar irradiation, manipulations of the 
total energy equation (\ref{eq:tmvisc}) imply that 
\begin{equation}
T_{\rm e} = \left(\frac{3GM_\star \dot{M}}{8\pi\sigma} \right)^{1/4} r^{-3/4} \mbox{   ,   }
\end{equation}
which can then be used to deduce $h$, as usual:
\begin{eqnarray}
\frac{\overline{h}}{1\mbox{AU}} &=& \left( \frac{3}{2}\frac{L_{\rm acc}}{L_\star}\right)^{\frac{n+\beta+1}{8n+8}} \left( \frac{4\theta^2 I(n)}{3\kappa_{\rm V} \overline{\Sigma} \tilde{Z}}\right)^{-\frac{1}{2n+2}} \left(\frac{\mu G M_\star}{R_{\rm g} R_\star T_\star}\right)^{-1/2} \nonumber \\ &\cdot & \left(\frac{R_\star}{1 \mbox{AU}}\right)^{\frac{3\beta-n-1}{8n+8}} \mbox{   ,   } \nonumber \\
\frac{q+3}{2} &=& \frac{9n-3\beta+9-4s}{8(n+1)} \mbox{   .   }
\end{eqnarray}
where we define here $L_{\rm acc} = G M_\star \dot{M}/R_\star$. In the case of steady accretion, we can reduce this further to yield
\begin{eqnarray}
\frac{\overline{h}}{1\mbox{AU}} &=& \left( \frac{3}{2}\frac{L_{\rm acc}}{L_\star}\right)^{\frac{n+\beta+1}{8n+16}}  \left( \frac{4 \theta^2 I(n)}{\kappa_{\rm V} \tilde{Z}} \frac{\pi \alpha_{\rm t}}{\dot{M}} \sqrt{\gamma G M_\star R_\star} \right)^{-\frac{1}{2n+4}} \nonumber \\
&\cdot& \left( \frac{\mu}{R_{\rm g}}\frac{G M_\star}{R_\star T_\star}\right)^{-\frac{n+1}{2n+4}} \left(\frac{R_\star}{1 \mbox{AU}}\right)^{\frac{3\beta-n+1}{8n+16}} \mbox{   ,   } \nonumber \\
\frac{q+3}{2} &=& \frac{9n-3\beta+15}{8n+16}\mbox{   .   }
\end{eqnarray}
With the adopted fiducial values for $n$ and $\beta$, we find that the
viscously heated regime is the only one that is not
flaring: indeed, in this case $q = -1.125$ and $s = 0.375$. However,
flaring is not self-consistently required towards the determination of
the disk structure in this region, so the formula given above is in
fact valid.  One may wonder, however, whether the viscously heated
regions could shadow the innermost radiatively heated regions, as
proposed by Dullemond \& Dominik (2004) for HAeBe stars. We do not
believe that this is the case for classical T Tauri stars: it should
first be noted that the disk is still very nearly flat ($q$ is very
close to -1), and secondly that the radial extent of the viscously
heated region is small (which is not the case for HAeBe stars).

\section{Position of the transition radii}
In this section, we derive analytical expressions for the position for
the transition radii across the various regions for the specific case
where the disk undergoes steady-state accretion. For
simplicity we give the formula for the specific case where $\beta = 1$ 
and under the assumption of steady-state accretion,
although they can also be easily derived in the general case. 
In practise, the transition radii are most easily computed numerically.

\subsection{Transition from optically thin to marginally opaque}

This transition occurs when $\tau_{\rm tot}^s = 1$, or equivalently
when
\begin{equation}
\frac{\dot{M} r^{-s} }{3\pi \alpha_{\rm t} \sqrt{\gamma
G M_\star} \overline{h}^2} \frac{\kappa_{\rm V} \tilde{Z} }{2} 
\left(\frac{R_\star^2}{r^2}\right)^{1/5} = 1 \mbox{   ,   }
\end{equation}
where $s$ and $\overline{h}$ are derived from equation
(\ref{eq:qthin}). Solving for the transition
radius $r^{\rm t\rightarrow m}$ yields
\begin{equation}
r^{\rm t\rightarrow m}_{\rm AU} = \left[ \frac{\tilde{Z} \kappa_{\rm V}}{2}
\frac{\dot{M} }{3\pi \alpha_{\rm t} \sqrt{\gamma G M_\star R_\star}}\right]^{\frac{90}{133}}  \left[
\frac{1}{C}\frac{45}{13} \left(\frac{\mu}{R_g}
\frac{GM_\star}{R_\star T_\star}\right)^{5} \right]^{\frac{20}{133}} \frac{R_\star}{1 \mbox{AU}}
\end{equation}
We see that, as expected, this transition moves inwards as the disks
evolve and their mass accretion rate decreases.

\subsection{Transition from marginally opaque to optically thick}

This transition occurs when $\tau_{\rm tot}^m = \tilde{Z} \Sigma \kappa_{\rm m} /2$
reaches 2/3, or equivalently, when the RHS of (\ref{eq:nasty}) reaches
unity. In a steadily accreting disk, with $\beta = 1$, the
right-hand-side of equation (\ref{eq:nasty}) reduces to
\begin{equation}
\frac{2}{3} \sqrt{\frac{2}{\pi}} \frac{1}{\tilde{Z} \rho_{\rm m} h \kappa_{\rm V}}
\left(\frac{T_{\rm m}}{T_\star}\right)^{-1} \simeq \frac{2 \sqrt{2\pi
\gamma} K \alpha_{\rm t}}{\tilde{Z} \kappa_{\rm V} \dot{M}}  \frac{R_g
T_\star}{\mu \Omega_{\rm K}} \mbox{   ,   }
\end{equation}
where $K$ varies slowly between $\sqrt{2\pi}$ (in the nearly-optically
thin case) and $2 I(n) \theta$ (in the very optically thick
case). This implies that for a given accretion parameters $\dot{M}$
and $\alpha_{\rm t}$ and given stellar parameters $T_\star$ and
$M_\star$, the transition radius at which the disk changes from
optically thick to optically thin to its own radiation is entirely
determined, and equal to
\begin{equation}
r^{\rm m \rightarrow o}_{\rm AU} = \left[ \frac{\tilde{Z} \kappa_{\rm V}}{2} \frac{\dot{M}}{2\pi \alpha_{\rm t}}  \sqrt{\frac{G M_\star}{\gamma R_\star^3} } \frac{\mu} {R_g T_\star} \right]^{2/3} \frac{R_\star}{1 \mbox{AU}} \mbox{   .   }
\end{equation}
Again, as expected, this transition moves inwards as the disks evolve
and their mass accretion rate decreases. It is interesting to note,
however, that to a good degree of approximation both $r_{\rm
t\rightarrow m}$ and $r_{\rm m\rightarrow o}$ are proportional to
$\dot{M}^{2/3}$, which implies that the ratio $r^{\rm t\rightarrow
m}/r^{\rm m\rightarrow o}$ is independent of mass accretion rate.

\subsection{Position of the snow line}

The outer edge of the snow line is located where $T_{\rm m}$ first reaches
160K. This transition can either happen in the optically thin
region or in the optically thick region. In both cases, its position
is given by the equation
\begin{equation}
r^{\rm snow}_{\rm AU} = \left(\frac{160 R_{\rm g}}{\overline{h}^2 \mu}\frac{(1\mbox{ AU})^3}{G
M_\star}\right)^{1/q} \mbox{   ,   }
\end{equation}
where $\overline{h}$ and $q$ are region-dependent. In the optically thick 
case, (for $\beta = 1$, and $n$ and $\theta$ given by equation 
(\ref{eq:z00})) we find that 
\begin{equation}
r^{\rm snow}_{\rm AU} \propto \dot{M}^{1/3}
\end{equation}
so the snow line moves inwards as the mass accretion
rate decreases. Since this migration rate is slower than the rate at which 
the optically thin region expands, there must be a point where the 
snow line becomes optically thin. When this happens, its location moves 
outward again to a radius which is a function only of the stellar parameters 
(with a weak dependence on the disk structure through $C$). When $\beta = 1$,
\begin{equation}
r^{\rm snow}_{\rm AU} = \left( \frac{160 R_{\rm g}}{T_\star}\right)^{-\frac{45}{19}} \left(\frac{45}{13C}\right)^{\frac{-10}{19}}
\left(\frac{\mu}{R_g} \frac{GM_\star}{R_\star T_\star}\right)^{-\frac{5}{19}} \frac{R_\star}{1\mbox{AU}}\mbox{   .   }
\label{eq:snowthin}
\end{equation}

The smallest possible radius attained by the snow line in the course of the disk evolution can  be estimated simply by finding the position of the snowline while it is located in the weakly opaque region. We find that 
\begin{equation}
\min\left(r^{\rm snow}_{\rm AU}\right) \simeq 0.6 {\rm AU} \left(\frac{L_\star}{L_\odot}\right)^{2/3} \left(\frac{M_\star}{M_\odot}\right)^{-1/3} 
\end{equation}
This can be considered as the minimum radius at which a significant amount of vapor-phase water can be found in protoplanetary disks in the course of their evolution. Most importantly, note that this value is independent of the value of $\kappa_{\rm V}$ selected, and of the nature of the polytropic solution selected: it is uniquely determined by the properties of the central star. However, the value of $\dot{M}$ for which this minimal position is achieved does depend on these disk properties.


In Section \ref{sec:discuss}, we discuss in more detail the evolution
of the position of the snow line as the disk evolves, and its
consequences for planet formation.

\subsection{Transition from radiatively heated to viscously heated}

This transition occurs when viscous heating and radiative heating are of the same order of magnitude. In a steady-state accretion approximation, we deduce the transition radius from equating
\begin{equation}
\frac{3}{4\pi} \dot{M} \Omega_{\rm K}^2 = 2\frac{A_{\rm s}}{2} \frac{L_\star}{4\pi r^2} \mbox{   ,   }
\end{equation}
to be
\begin{equation}
r^{\rm visc}_{\rm AU} = \left( \frac{3 L_{\rm acc}}{S L_\star} \frac{R_\star}{
\overline{h}} \frac{2}{q+1} \right)^{\frac{2}{q+3}}\mbox{ , }
\end{equation}
where the constant $S$ is either $\theta$ if the transition radius
occurs in strongly opaque regions, or $C$ if it is located in
optically thin, marginally or weakly opaque regions. The exact values
of $\overline{h}$ and $q$ also depend on the radial region considered.

\section{Examples of star/disk systems}
\label{sec:examples}

\subsection{Examples of protostellar disks around a weak-line T Tauri star}
\label{sec:ttauri}

We now apply our model to dusty disks around T Tauri stars. We defer
to later publications the discussion disks around the more massive
Herbig AeBe stars and around the low-mass brown dwarves. Direct
comparison between our model and the predictions of the CG97 model are
presented in the discussion section.

\begin{figure}
\epsscale{2.0}
\plotone{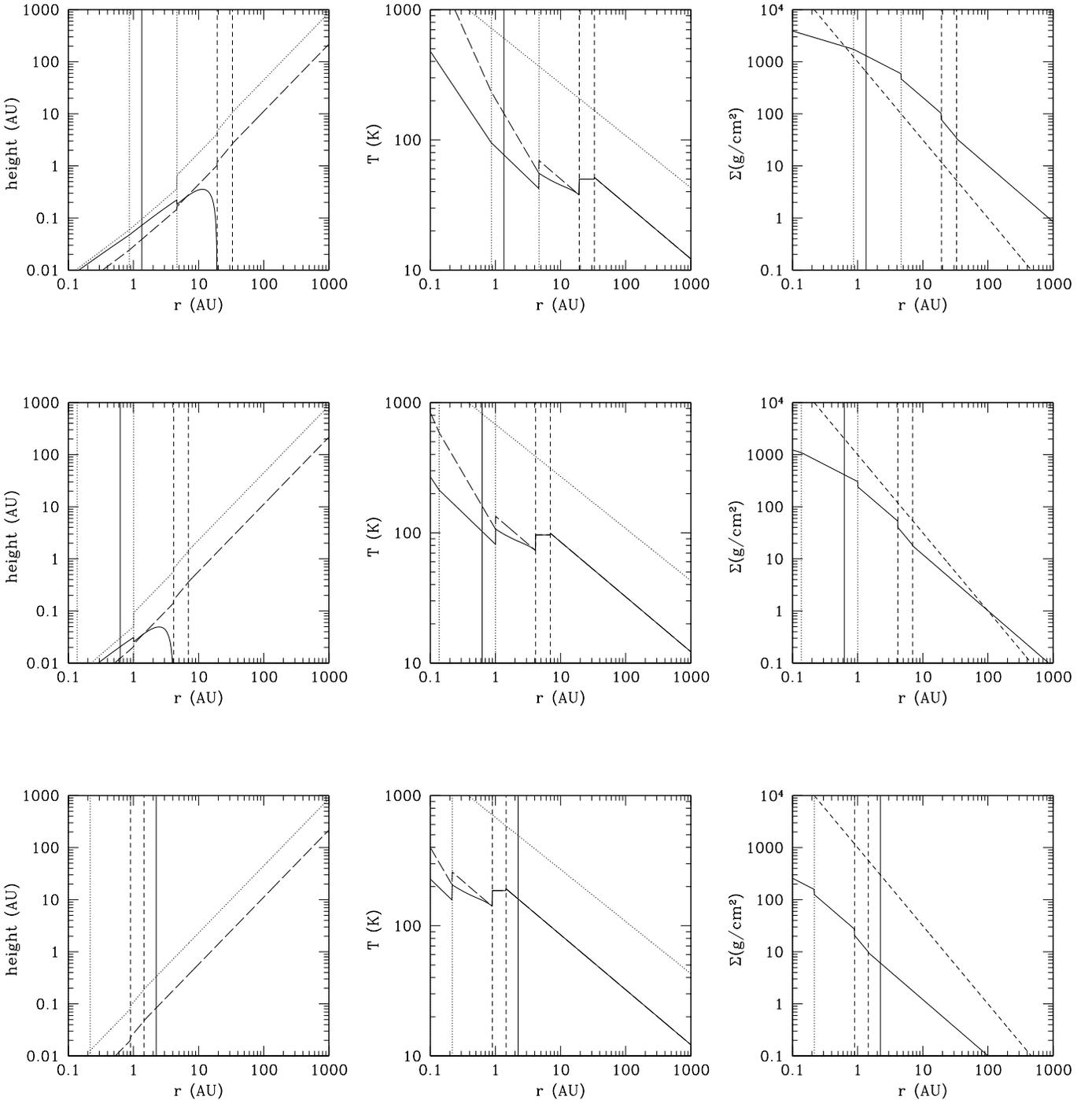}
\caption{Model prediction for a T Tauri star disk with accretion rates
(from top to bottom) of $\dot{M} = 10^{-8}, \dot{M} = 10^{-9}$, and $
\dot{M} = 10^{-10} M_\odot/$yr. The left-side figure shows the various
important surfaces and heights, the central figure shows the relevant
temperature profiles, and the right-side figure shows the surface
density profiles. On the left, the dotted line marks the position of
the superheated layer $z_{\rm s}(r)$, the long-dashed line marks the disk
temperature scale height $h(r)$, and the solid line marks the position
of the photosphere $z_{\rm e}(r)$. On the center, the dotted line marks the
temperature of the superheated grains $T_{\rm s}$, the long-dashed line
marks the temperature of the mid-plane $T_{\rm m}$ and the solid line marks
the effective temperature $T_{\rm e}$. On the right, the solid line marks
our model prediction for $\Sigma(r)$ while the thick dashed line is
the standard MSN profile $\Sigma_{\rm MSN} = 1000 r_{\rm AU}^{-3/2}$g/cm$^2$. 
In all diagrams, the snow line for the mid-plane
temperature appears as a vertical solid line.  The vertical dashed
lines (when present, from left to right)  denote the physical
transitions from the optically thick to the marginally  optically
opaque region, and then to the optically thin region. The vertical
dotted lines marks the position of the artificial transitions between
asymptotic regimes: near the central star (and in these diagrams, on
the left of the snow line when present) the transition from viscously
heated to radiatively heated, and near the marginally opaque region,
the transition from the largely opaque region to the weakly opaque
regions. }
\label{fig:ttauri-dotm}
\end{figure}

For the purpose of computing the structure of the disk, we select the
following values for the global stellar and disk properties (except when explicitly mentioned):
\begin{eqnarray}
&& L_\star = 1 L_\odot \mbox{   ,   } M_\star = 1 M_\odot \mbox{   ,   } R_\star = R_\odot \mbox{   ,   }\nonumber \\
&& \tilde{Z} = 2 \mbox{   equivalent to   } Z = 0.02  \mbox{   ,   }\nonumber \\
&& \alpha_{\rm t} = 10^{-3}  \mbox{   ,   }  \nonumber \\
&& \mu = 2.4 \mbox{   ,   } \gamma = 1.4  \mbox{   ,   } \\
&& n = 2  \mbox{   ,   } \nonumber \\
&& \kappa_{\rm V} = 1 {\rm cm}^2/{\rm g} \mbox{   ,} \beta = 1 \nonumber \\
&& C = 4  \mbox{   ,   }
\end{eqnarray}
and vary the mass accretion rate from $\dot{M} = 10^{-8} M_\odot/$yr
to $\dot{M} = 10^{-10} M_\odot/$yr, to cover a range of gas
accretion rates from observed values for classical T Tauri stars
(Hartmann 2001) down to low values presumably corresponding to a
protostellar-debris disk transition.  The results are shown in figure
\ref{fig:ttauri-dotm}.

\begin{figure}
\epsscale{2.0}
\plotone{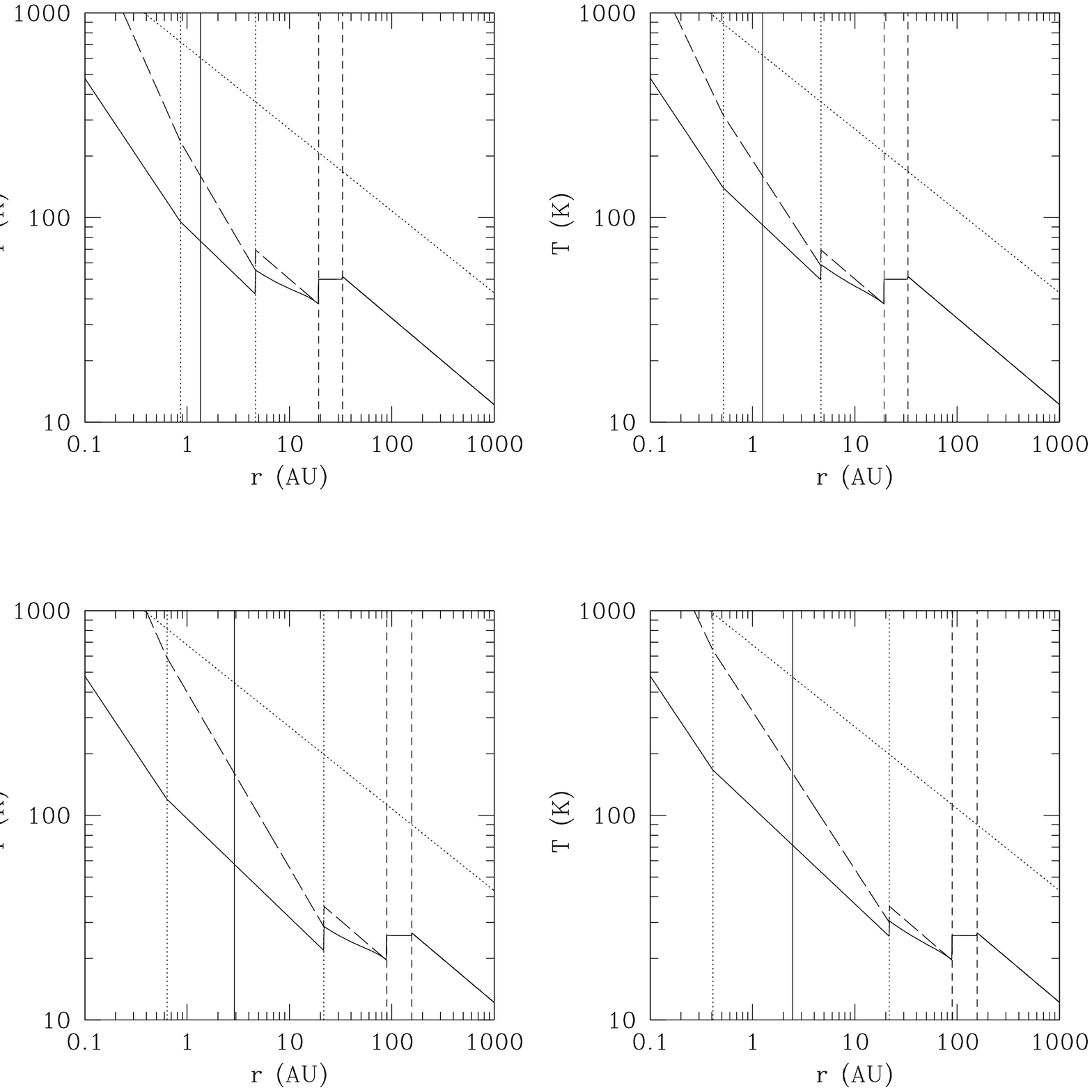}
\caption{Comparison of the model predictions for the temperature profiles of a disk with $\dot{M} = 10^{-8} M_{\odot}$/yr around a solar-type star, for two values of $\kappa_{\rm V}$ (from top to bottom $\kappa_{\rm V} = 1$cm$^2$/g and $\kappa_{\rm V} = 10$cm$^2$/g) and two values of the polytropic index $n$ (from left to right $n=2$ and $n=3$). The four figures are otherwise generated using the same conventions as in Figure \ref{fig:ttauri-dotm}. Note that the choice of the polytropic index $n$ has only a small qualitative influence on the solution in the optically thick regions of the disk (and naturally no influence on the optically thin regions), while the opacity $\kappa_{\rm V}$ influences strongly the position of the transition between optically thin and optically thick regions of the disk. }
\label{fig:ttauri-compare}
\end{figure}

Comparisons of the solutions for a particular value of $\dot{M}$
for two values of the polytropic index $n$ and the grain opacity
$\kappa_{\rm V}$ are shown in Figure \ref{fig:ttauri-compare}. We note
that while the choice of the polytropic index only has a small effect
on the radial structure of the disk, the choice of the opacity
naturally influences the transition between the optically thick and
optically thin regions of the disk.

For each accretion rate, three figures are shown: 1) the various
relevant surfaces ($h(r)$, $z_{\rm s}(r)$ and $z_{\rm e}(r)$), 2) the relevant 
temperature profiles ($T_{\rm m}(r)$, $T_{\rm s}(r)$ and $T_{\rm e}(r)$) and 3) the 
surface density profile. In each case, 
the superheated grains position and temperature
are shown as a dotted line. Note that $T_{\rm s}$ depends only on the nature
of the central star: the profiles obtained in all three values of $\dot{M}$ 
for the
particular T Tauri star chosen are identical. The position and
temperature of the photosphere are shown as a solid line. The
mid-plane temperature and its associated disk scale height $h$ are
shown as long-dashed lines.

The three regions (from right to left, optically thin, marginally
optically thin and optically thick -- the latter subdivided into weakly opaque,
 strongly opaque and viscously heated regions) are 
clearly visible in each figure. The transition
radii from optically thin to marginally opaque, and from
marginal to optically thick are genuine physical transitions, and are
marked by vertical dashed lines. We observe that the transition from 
an optically thin to a marginal region is associated with a fairly 
continuous temperature profile, while that from marginal to optically thick 
has an important temperature discontinuity. Note that in a real disk 
the temperature ``jump'' would be
more diffuse (as a result of radial thermal and viscous diffusion processes
unaccounted for in our 1+1D model), and could span a radial extent of up to
one scale height. The artificial transitions and associated discontinuities 
observed in the opaque region, (from left to right) between radiatively heated 
and viscously heated regions, and between the weakly opaque and strongly 
opaque regions, are mathematical artifact of our crude patching procedure 
between various asymptotic
approximations (in the former case, for the dominant heating mechanism and in 
the latter for equation (\ref{eq:nasty})); they are represented by 
dotted vertical lines. One can estimate the true profile by 
smoothly matching one region to
the next across these dotted lines by eye, or, should one desire a more
accurate solution, actually solve the problem numerically. 

Finally, the vertical solid line represents the snow line for the
mid-plane temperature. Across this line, the equations presented here
are only qualitatively valid, since we neglect all effects associated with the sublimation of the ices. We shall describe in detail a new
snow-line model in a following paper. Note that the temperature of the
superheated grains reaches 160K at a far greater radius than the marked
(mid-plane) snow line. The sublimation of the icy particles in 
the superheated layer  results in a drop in $\tilde{Z}$ above $z_{\rm s}$, 
but as shown in equation (\ref{eq:C(r)}),
this would only have a small effect on the value of $C(r)$. Hence
in what follows we ignore the effect of the successive sublimation
lines on the structure of the superheated layer. 

Note that since the outer edge of the disk depends on a number of potentially 
important factors (initial radius, self-shadowing in the very optically
thin case, total mass, thin-disk approximation, etc...), we have not marked 
it on the figures presented. The reader should also bear in mind that the 
model has less predictive power in regions much beyond 100 AU, where the 
assumptions inherent to steady-state accretion are no longer satisfied; in 
these regions, the problem is best solved as an initial value problem.

Several notable qualitative features can be readily seen from these
T Tauri star models and are also found in disks with a wide range of
accretion rates, around stars with a wide range of stellar
types. 

Firstly, we observe the presence of a wide, radially isothermal region
(corresponding to the marginally opaque region). Although the
isothermal nature of this section of the disk was already inferred by
CG97, we find that it also has the remarkable property of being
notably hotter than the region immediately within, so that {\it the
overall photospheric temperature profile is not a monotonically
decreasing function of radius}.  The hotter temperature is a result of
the larger flaring angle of this region compared to the one within,
implying that a larger fraction of the stellar flux is absorbed
locally. The presence of a ``hot ring'' could have very interesting
consequences, both from an observational 
point of view and from the point of view of the dynamics of dust
grains in the disk. This is discussed in detail in 
\S\ref{sec:discuss}.

Secondly, although a more rigorous snow-line model will be needed for
more accurate quantitative discussions, 
it is already apparent in the models presented that the radial
extent of the snow line can be considerable when it 
lies within the optically thick region. By radial extent, we
imply the region between the radius at which the mid-plane temperature
becomes equal to the ice-sublimation temperature, to the radius at
which the photospheric temperature reaches the same value. For
example, our results indicate that for a standard accretion rate of
$10^{-8} M_{\odot}/$yr, the snow line extends from 0.4 AU (where $T_{\rm e}
= 160$K) to 1.35 AU (where $T_{\rm m} = 160$K). We also note that its extent
increases with mass accretion rate (see also Figure \ref{fig:snow}).
On the other hand, for low mass accretion rates (for a
1$M_\odot$, 1$L_\sun$ T Tauri star when $\dot{M} < 10^{-10}
M_{\odot}$/yr) we find that the snow line is located in the optically
thin region, with entirely different physical and dynamical
characteristics. The consequence of the transition from an optically
thick to an optically thin snow line will be discussed in more detail
in a forthcoming paper.

Thirdly, we note that the predicted overall surface density profile is
notably shallower than a Minimum Solar Nebula model. Only in the
marginally opaque region do we reproduce the standard MSN radial power
law scaling. This
will be discussed in more detail in \S7.

\section{Discussion of the results and observational prospects}
\label{sec:discuss}

Our new disk model is based on that proposed by CG97, 
but it contains two significant
additions. In our analysis, a polytropic assumption for the disk
in optically thick regions enables us to study the largely
non-isothermal structure of typical inner disks, while retaining the
analytical nature of the mathematical solutions. We have also
incorporated the contribution to heating due to viscous dissipation which
necessarily accompanies mass accretion. In addition,
unlike CG97 we do not assume a MSN surface density profile. We can
therefore calculate self-consistently all disk quantities as a
function of radius assuming steady-state accretion or, as we shall
consider in a future investigation, solve numerically for the temporal
evolution of the disk as an initial value problem, 
using the mass accretion rates derived in the model

In this section, we begin by a direct comparison of our model
predictions with those of CG97 using their chosen
fiducial T Tauri star characteristics. Then, we critically discuss the
model assumptions and finally, the model predictions in the light of
recent observational and theoretical advances.

\subsection{Comparison with the Chiang \& Goldreich model}

\begin{figure}
\epsscale{2}
\plottwo{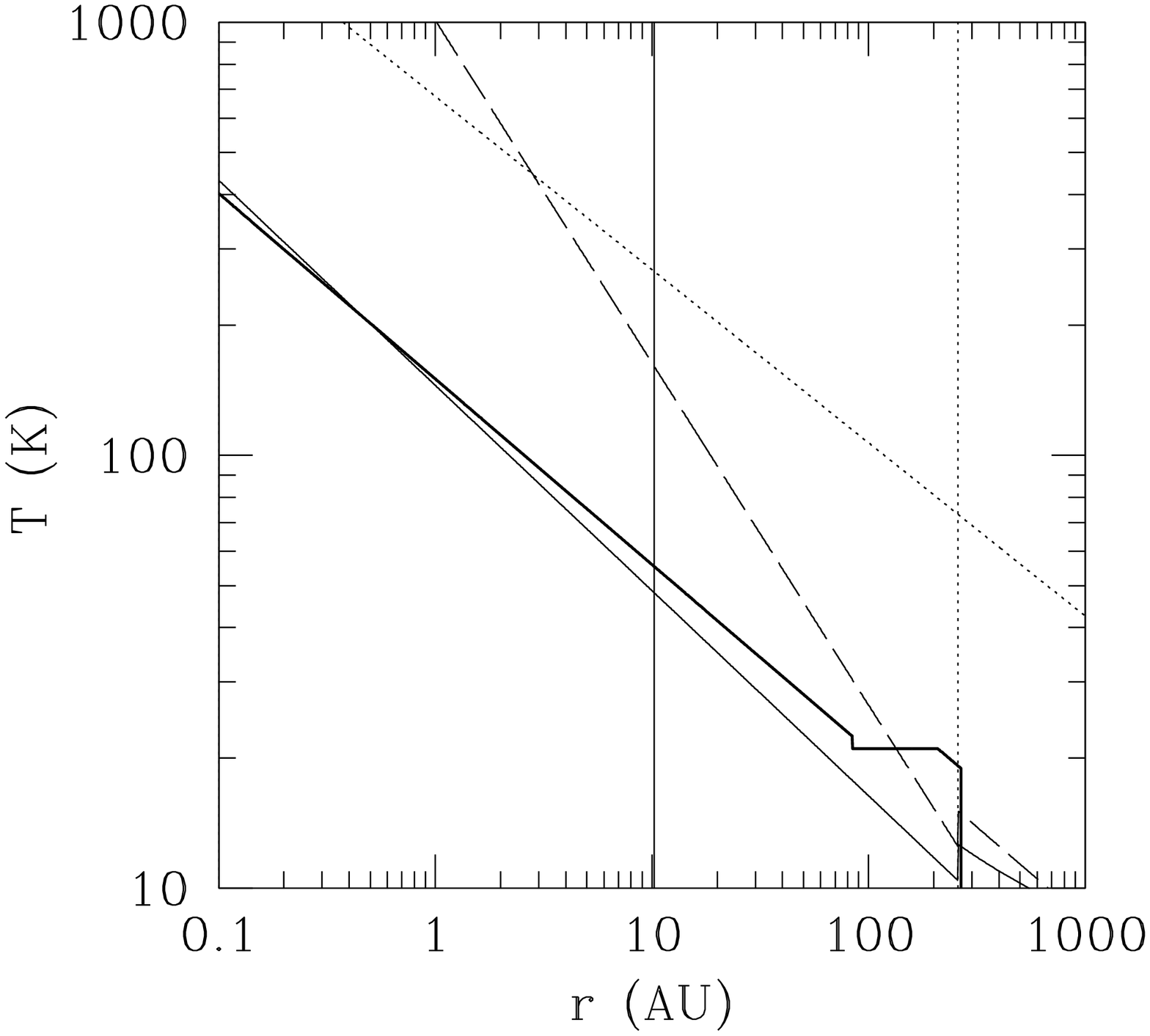}{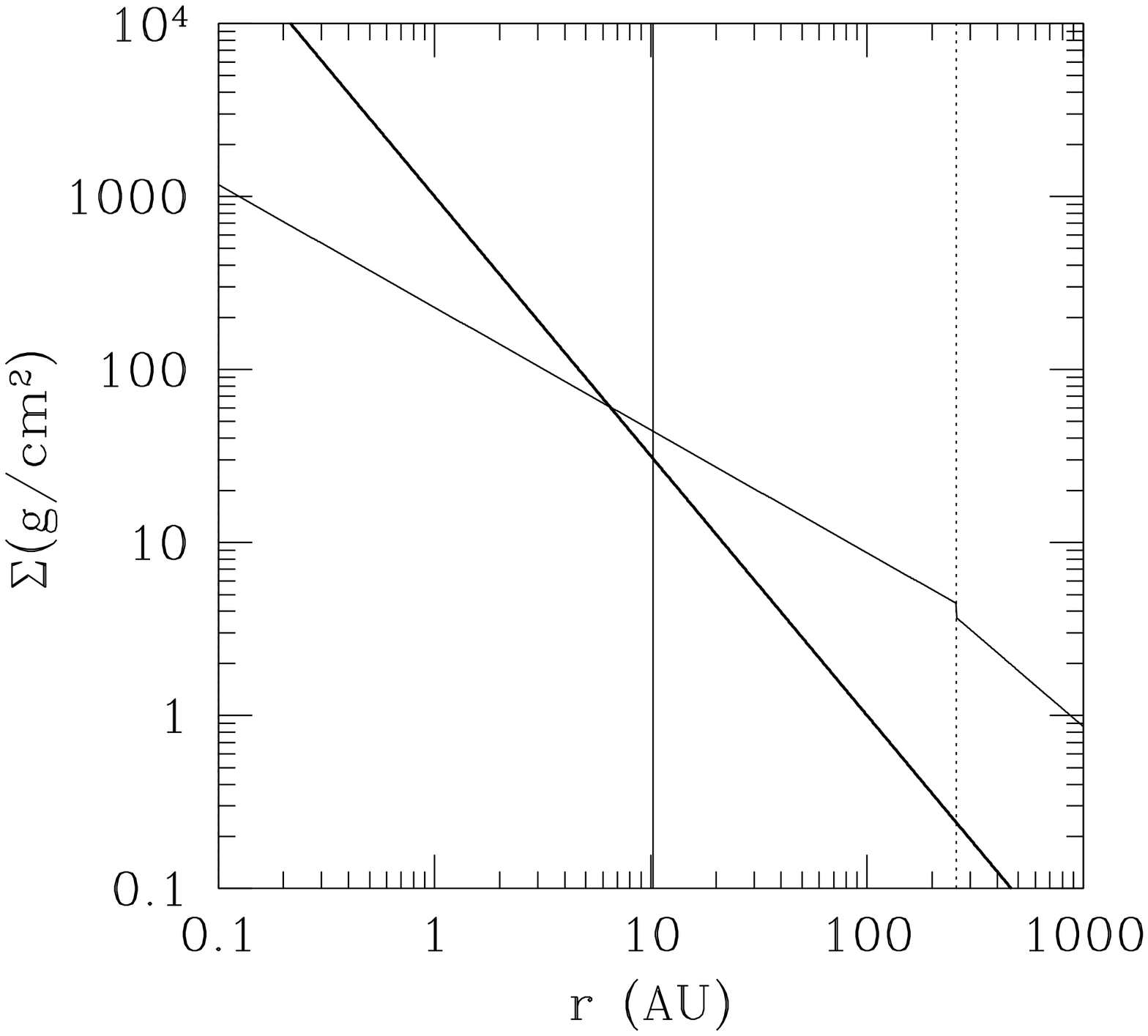}
\caption{Direct comparison between our model predictions (thin lines,
line-style coding as in Figure \ref{fig:ttauri-dotm}) and the CG97
model predictions (thick line) for the temperature and surface density
in a disk around a typical T Tauri stars with mass $0.5 M_\odot$,
radius $2.5 R_\odot$ and effective temperature 4000 K. Note in
particular the lower effective temperature of our model, the much
higher mid-plane temperature, and the higher surface density in the
outer regions. The predictions for the temperature of the superheated
grains are naturally identical.}
\label{fig:chiang}
\end{figure}

We compare our model predictions for the temperature profile in a
protostellar disk around a $0.5 M_\odot$, $2.5 R_\odot$ T Tauri star
to the formula given by CG97 based on their isothermal model. A
straightforward comparison is impossible since the mass accretion rate
in their model varies between about $5 \times 10^{-7}$ to about $2
\times 10^{-8} M_\odot$/yr (assuming the same value of $\alpha_{\rm
t}$ in both models), as a result of their assumed surface density
profile. On the other hand, we need to fix $\dot{M}$ so we choose a
fiducial value of $\dot{M} = 10^{-8} M_\odot/$yr for the purpose of
the comparison. In addition, for this simulation only we change the 
opacity $\kappa_{\rm V}$ to the value of 400cm$^2$/g used by CG97. 
The results are shown in Figure \ref{fig:chiang}.

We note first that regardless of the inherent difficulty in comparing
models with different assumptions, the predictions for the
photospheric temperatures do, at a first glance, match fairly
well. This agreement is expected since the only difference in
the total energy budget in the optically thick, radiatively heated
regions of the disk arises from slightly different predicted flaring
angles in the two approaches. However, beyond this overall
match, we note several points of discrepancy.

Firstly, the photospheric temperature in the optically thick inner
nebula in our model is in fact systematically lower than theirs
by 10-20\%, except very near the central star. This difference solely
arises from the lower flaring angle predicted when heating is
taken into account in the disk mid-plane.

Secondly, the position of the transition between optically thick and
optically thin regions of the disk is closer to the central star in
their model than in ours. This mis-match is caused by our
largely different prediction for the surface density profile of the
disk, which implies that the outer regions have higher overall
mass content (and optical depth) than in a standard MSN model for the
mass accretion rate considered.

Finally, note that their isothermal model predicts a snow line located
around 1AU, while taking into account the non-isothermal nature of the
optically thick region implies that the snow line extends as far out
as 10 AU for this particular T Tauri star. Indeed, at 1AU for instance our mid-plane temperature is $\sim 10^3$K, and exceeds the photospheric temperature by a factor of five. Neglecting this effect can lead to a gross mis-calculation of the position of the snow-line.

\subsection{The validity of the model assumptions}

The three major assumptions of the proposed model are the following:
viscous dissipation is modeled by an ad-hoc $\alpha-$prescription,
the disk stratification is modeled as a polytrope in optically
thick regions and finally the dust grains are assumed to be uniformly
mixed with the gas at all heights and radii.

Turbulent viscous dissipation affects the disk in two ways: it
promotes the outward transfer of angular momentum enabling gas accretion
onto the central star, and generates heat preferentially in the
mid-plane of the disk. 
For modest accretion rates, the viscous contribution to the
total energy budget of the disk at any particular radius is
significant only well within the snow-line. However, it does
does play a role in driving a small temperature gradient between the
mid-plane and the photosphere in optically thick regions well outside of the snow-line.  By modeling the turbulent viscous dissipation using an
$\alpha-$prescription we implicitly assume that the mechanism driving
turbulence is robust, and independent of position within the disk.

While observations ubiquitously suggest that accretion is an ongoing process 
in protostellar disks (Hartmann 2001), and that it must be 
approximately quasi-steady (otherwise, strong accumulation of
matter near specific locations in the disk would be systematically
observed), the origin of turbulence still remains largely
unclear. Turbulent mixing cannot be of convective nature only, since
heating of the surface layers may suppress convection (Watanabe {\it
et al.} 1990). Mixing by a magneto-rotational instability (Balbus \&
Hawley 1990) is difficult to sustain in regions where the ionization
fraction of the gas is too low (the dead-zones).  Although surface
layers of the disk are always potentially subject to the MRI since
they are exposed to a significant ionizing flux both from cosmic rays
and from the central star (Gammie 1996), it is not clear that turbulent
motions driven high above the disk are strong enough to sustain a high
$\alpha-$value all the way down to the disk mid plane.  Finally,
turbulent mixing by a sub-critical instability of the background
Keplerian shear has also been proposed (Richard \& Zahn 1999, Dubrulle
{\it et al.} 2005), as well as shearing instabilities in the mid-plane
layer caused by the sedimentation of the dust particles (Goldreich \&
Ward 1973, Weidenschilling \& Cuzzi, 1993 Garaud \& Lin 2004, Gomez \&
Ostriker 2005), but while the former has not yet been formally proven to 
exist, the typical growth rates of the latter may be too long compared 
with the dynamical timescale to be effective in providing a reliable 
source of angular momentum mixing. 

To complicate matters, it is likely that 
the intrinsic disk stratification (both radially and
across the mid plane) may result in largely non-uniform
turbulent transport properties. Thus we recognize that the $\alpha-$
prescription is merely an ad-hoc assumption, necessary to our
simplified treatment, and now discuss to which extent the results we
obtain are model-dependent.

In our model, the $\alpha$-prescription is used
to calculate the surface density profile of disks under the
assumption of steady-state accretion. Consequently, if 
$\alpha$ varies significantly with radius, the
radial profiles obtained in the steady-state accretion assumption must
be modified accordingly, and the radial scaling laws obtained should
be revised entirely. 

The generation of heat by viscous dissipation is on the other hand 
subtly disguised in our model by the assumption of a polytropic stratification
for the gas/dust mixture. Thus the model is indirectly dependent on $\alpha$
through the polytropic index assumed. 

Finally, the last questionable assumption of the model involves the
fully mixed nature of the dust grains with the gas. Grains are fully
suspended in the gas only when their stopping time is much longer than
the turnover time of the turbulent eddies (which no more than one
order of magnitude larger than the orbital time). For the smallest
grains (typically, 0.1$\mu$m) which are responsible for most of the
absorption of the stellar light within the superheated layer, this
approximation is largely satisfactory at all radii. Thus we do
not expect the position and structure of the superheated layer to be
affected by this simplified approximation: significant sedimentation
of larger grains (mm-size) will leave the superheated layer
unaffected.  On the other hand, the disappearance of the smallest
grains resulting from the {\it overall} growth may result in a change
in $C(r) = z_{\rm s}/h$ by a factor of order unity. This possibility should
be taken into account self-consistently in future models. With respect
to the mid-plane and radial structure, grain growth results in a change
in $\kappa_{\rm V}$ and $\beta$, which will affect the radial
temperature distribution (and therefore the spectral slope) in all
regions of the disk. Sedimentation doesn't affect the re-processing of
light in the optically thin regions, but may have a detectable effect
on the position of the photosphere (and therefore the radial profile)
in the optically thin regions.

\subsection{Discussion of the results and their observational implications}

\noindent {\it Disk geometry}:\\ As in CH97, we find that the stellar
radiation is primarily absorbed by a flaring layer ($\alpha > 0$) of
superheated grains in the radiatively heated regions, including in the
newly studied case of optically thick, non isothermal disks.  Regions
where viscous dissipation is the principal source of energy are, on
the other hand, found to be non-flaring ($\alpha < 0$) although their
overall geometry is very close to flat.

From the central regions to the outer regions of the disk we find that
the flaring angle gradually increases outward.  The viscously heated,
and radiatively heated, strongly opaque regions of the disk are close
to a flat geometry ($\alpha = 0$), with the temperature index $q$
increasing from $-1.125$ to $-0.85$ (corresponding to $\alpha \simeq
-0.0625 h/r$ and $\alpha \simeq 0.075 h/r$ respectively). Thus we note
that a flat disk geometry does not necessarily imply any grain
sedimentation, and it can also straightforwardly be reproduced by a
appropriately modeled optically thick disk.

The marginally opaque transition region (from optically
thick to optically thin disks) is accompanied by a very
sharp, local increase in the flaring angle with $\alpha = 0.5 h/r$.
This transition is therefore accompanied by a sharp local increase in
the temperature, reaching values up
to 40\% larger than in the optically thick region immediately within.

Self-consistent calculations of the height of the superheated layer in
the optically thin regions shows that the flaring geometry persists
all the way out to several hundred AUs, albeit with a slightly smaller
flaring angle ($\alpha \simeq 0.3 h/r$). 

Self-shadowing of the
outermost regions may occur, as discussed earlier, when $\alpha$
exceeds a fraction of the total visual optical depth, unless other
mechanisms truncating the disk occur first. This sharp cut
off in the reprocessed radiation could appear as an outer edge of the
disk.
\bigskip

\noindent {\it Disk SEDs and surface density distribution}\\ 
The MIR and FIR photons are emitted by both grains at photosphere of
the inner opaque regions where $T_{\rm e} < 200$K and by the superheated
grains with $T_\star < 200$K at much larger radii.  Around classical T
Tauri stars where $\dot M \sim 10^{-7} M_\odot$ yr$^{-1}$, the radii
where $T_{\rm e}$ and $T_\star$ decrease below 200K are $0.7$ and 30 AU
respectively.  Around weak-line T Tauri with $\dot M \sim 10^{-9}
M_\odot$ yr$^{-1}$ the radii where they fall below $200K$ are 0.07 and
30 AU respectively.  Although the area filling factor $A_{\rm s} \sim
\alpha$ of the superheated grains with $T_\star < 200$K is well below
unity, their total emitting area is much larger than the opaque inner
region with $T_{\rm e} < 200$K.  Consequently, the superheated grains
contribute to a large fraction of the MIR and FIR radiation and the
SED in this wavelength range is essentially flat (CG97).

The magnitude of $T_{\rm e}$ of the opaque regions is generally too
high for the emerging photons to contribute significantly to the
sub-mm radiation.  These photons are emitted by much-cooler and
much-larger (mm to cm size) grains near the mid plane of marginally-
transparent and optically-thin regions in the outer disk.  At very
large disk radii, the superheated grains also contributes to these
long wave radiation.

A robust feature of our model is the prediction of an overall disk
surface density much shallower than the standard MSN; this prediction
is associated with the assumption of steady-state accretion.  More
precisely, we find that the opaque inner disk regions satisfy roughly $\Sigma
\propto r^{-0.4}$ and $r^{-0.7}$ in the viscously and radiatively
heated regions respectively.  The marginally opaque disk regions,
however, are radially isothermal which implies a surface density
profile varying as the MSN profile (Hayashi {\it et al}. 1985) under
the steady-state accretion approximation.  This is particularly
interesting since these intermediate regions can have a large radial
extent (between 100-200 AU and 4-7 AU respectively as
$\dot{M}$ decreases from $10^{-7}M_{\odot}$/yr to $10^{-9}M_\odot$/yr). 
Finally, in the optically thin regions we find
$\Sigma \propto r^{-1.08}$. Around some T Tauri stars, the outer
regions have been resolved by mm interferometer observations.  Our
prediction on the surface density distribution in this region is very
similar to that inferred from the
mm maps of resolved protostellar disks around T Tauri stars (Mundy
{\it et al.} 2000, Looney {\it et al.} 2003). The disk mass determination 
based on the
assumption that these regions are optically thin (Beckwith \& Sargent
1991) is fully justified.
\bigskip
 
\begin{figure}
\epsscale{1}
\plotone{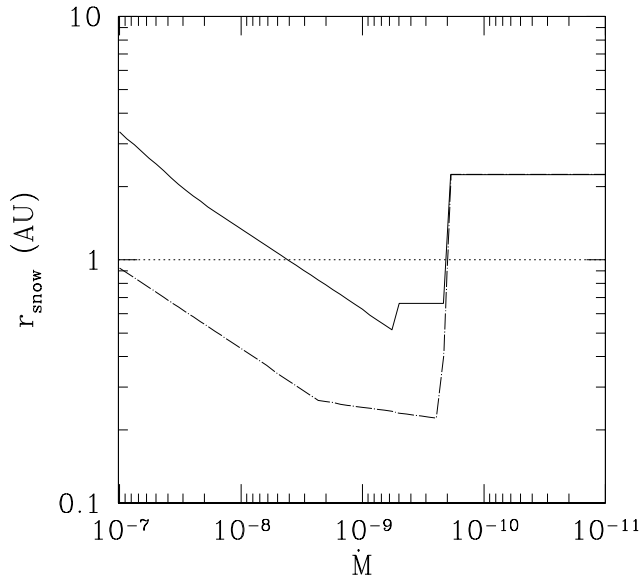}
\caption{Variation of the position of the snow line as a function of
mass accretion rate for a 1$M_{\odot}$, 1$L_{\odot}$ T Tauri star. The
solid line represents the location where the mid-plane temperature
$T_{\rm m}=160K$ whereas the dash-dotted line represents that 
where the effective
temperature $T_{\rm e}=160K$.  Between these two locations, ice can retain
solid phase in a fraction of the disk near its surface.  Since a large
fraction of the grains are located near the mid plane, along with the
turbulent gas, the ice line for the mid-plane temperature is more
relevant for determining the phase of most water molecules.  The slow
decrease in the position of the snow line (for $T_{\rm m}$) as $\dot{M}$
initially decreases corresponds to the phase in which the snow line is
located in the strongly opaque region. The first plateau in
this profile arises when the snow line moves into the weakly opaque
 region. Therefore the observed jump is an artifact
of the mathematical asymptotics, and should be interpreted instead as
a smooth flattening of the profile. The second plateau, however,
corresponds to the genuine physical transition of the snow line moving
from the optically thick to the optically thin region as the disk
depletes.}
\label{fig:snow}
\end{figure}

\noindent {\it Snow line position, structure and
history}\\ As seen in Figure \ref{fig:ttauri-dotm}, 
the snow line around $1 M_\odot$ classical T
Tauri stars shrinks from a few AU to about 0.6AU as $\dot
M$ decreases from $10^{-7}$ to $10^{-10} M_\odot$yr$^{-1}$, then
re-expands to slightly over 2.2 AU.  This interesting non-monotonous
variation of the snow line position with mass accretion rate (and
therefore with time) is illustrated in Figure \ref{fig:snow}.

For relatively high values of the mass accretion rate, 
we confirm that the snow
line is located in the strongly opaque region, and therefore has a
significant radial extent. For 
$\dot M \sim 2 \times 10^{-7} M_\odot$ y$^{-1}$ the snow line is located
near the present-day radius of Jupiter; it may have been critical in
promoting the emergence of the first gas giant planet at this location
in the solar system (Ida \& Lin 2004).

During the final stage of disk depletion,
the snow line moves into the optically thin region. For a
1$M_{\odot}$, 1$L_{\odot}$ T Tauri star this transition occurs when
$\dot{M}$ drops below $2 \times 10^{-10} M_\odot$ yr$^{-1}$. Note that
the sharpness of the transition observed in Fig \ref{fig:snow} may be
an artefact of the simplistic snow line model used here; we will study
the physics of the snow line and the consequences of the transition
from optically thick to optically thick in more detail in 
a subsequent paper.

Most interestingly, we find that the position of the snow line drops
down to a radius of about 0.6 AU before expanding out again. This
location is comparable to the inner boundary of the habitable zone
around solar-type stars (Kasting {\it et al.} 1993). The discrepancy
between our results and those obtained by
Lecar {\it et al.} (2006) for instance arises from our self-consistent 
calculation of the flaring angle in 
the optically thick regions of the disk, combined with the assumption of
steady-state accretion. The consequences of this finding are
potentially far reaching.  When the disk mass accretion rate has
declined to about $5 \times 10^{-9} M_\odot$ yr$^{-1}$, the snow line crosses
the Earth's orbit and steadily continues shrinking; at this point, the
local total surface density of the disk is about 400 g/cm$^2$, so that
in a ring of with a width $x r_{\rm hill}$ around the Earth's orbit
one may find about $2 \times 10^{-2} x M_\earth$ of condensed water in the form
of icy particles.  About half this amount in ices is also found
around the orbit of Venus when the snow line crosses its orbit. 
This amount largely exceeds the current total water content of Earth and Venus, and could provide an alternative explanation for their hydration:
the possibility of a late stage accretion of ice grains onto large
protoplanetary embryos {\it in situ} has the main advantage of
providing a simple deterministic explanation for the measured D/H
abundance ratio (Lunine 2005) of the oceans' water. 

Finally, we also note that T Tauri stars younger than 1 Myr undergo FU Orionis
outbursts (Herbig 1977, Hartmann \& Keyon 1985).  During these
outbursts, the central luminosity increases above $100 L_\odot$
(Hartmann 2001) and the snow line can near-instantly expand well 
beyond 100AU, then retreat back to its initial position in a
short period of time ($\sim$ 100 yr). This episodic relocation of the
snow line will clearly be very disruptive to the growth of any icy
body in most regions of the disk within the first $10^5$ yr 
of the disk's evolution. This phenomenon may be observable through 
the coexistence of water vapor and ice at large radii in very young 
protostellar disks.\\
\\
We thank Drs T. Guillot, K. Kretke, C. Lada, S. Mohanty, G. Ogilvie,
J. Pringle and A. Youdin for useful
conversation.  This work was supported in part by NASA
(NAGS5-11779,NNG06-GF45G, NNG04G-191G), JPL (1270927), NSF
(AST-0507424), and the California Space Institute.

\section*{Appendix}

This Appendix contains some of the mathematical tools and supplements
used towards the derivation of many of the equations from this
paper. 
\bigskip

\noindent {\it 1. Basic asymptotic integration method:} \\ Suppose
we attempt to integrate a function $e^{-f(x)}$ within the interval
$[a,b]$, when $f(x)$ has a minimum at $x=a$. Then
\begin{eqnarray}
\int_a^b e^{-f(x)} \dd x &\simeq& \int_a^b e^{-f(a)-(x-a)f'(a)} \dd x
= e^{-f(a)} \int_0^{b-a} e^{-u f'(a)} \dd u \nonumber \\ &=&
\frac{e^{-f(a)}}{f'(a)} \left(1-e^{-(b-a)f'(a)}\right) \simeq
\frac{e^{-f(a)}}{f'(a)} \mbox{   ,   }
\end{eqnarray}
in the limit where $(b-a)f'(a) \gg 1$. Naturally, the curvature terms
neglected in the expansion of $f(x)$ near $x=a$ also set another
condition on the applicability of the approximation that must be
checked for self-consistency.
 
When $b \rightarrow +\infty$, this approximation can be used for
instance to recover the well-known approximation to the complementary
error function as $x \rightarrow +\infty$:
\begin{equation}
{\rm erfc}(x) = \frac{2}{\sqrt{\pi}} \int_x^{+\infty} e^{-u^2} \dd u \simeq \frac{1}{\sqrt{\pi}} \frac{e^{-x^2}}{x}  \mbox{   .   }
\label{eq:erfc}
\end{equation}
This approximation is used in the derivation of equation
(\ref{eq:tauvert}).
\bigskip 

\noindent {\it 2. Position of the superheated layer} \\
The derivation of equation (\ref{eq:tau0}) is straightforward using
this method. Indeed,
\begin{equation}
\int_0^r \tilde{Z} \kappa_{\rm V} \rho(z',r') \dd {\bf l} = \kappa_{\rm V}
\int_0^r \rho_{\rm m}(r') e^{- \frac{z_{\rm s}^2}{r^2}\frac{r'^2}{2h^2(r')}} \dd
r' \mbox{   ,   }
\end{equation}
so that we can use the above method with 
\begin{eqnarray}
&& a = 0 \mbox{   ,   } b = r \\
&& f(r') = \frac{z_{\rm s}^2}{r^2}\frac{r'^2}{2h^2(r')} - \ln (\tilde{Z} \rho_{\rm m}(r')) \mbox{   ,   } 
\end{eqnarray}
(note that we thereby consider the possibility of $\tilde{Z}$ having a slow
variation with $r$). Here, the function $f(r')$ has a minimum at $r'=b$
provided the disk is flaring (i.e. provided $h/r$ is an increasing
function of $r$). In that case, the formula used is
\begin{equation}
\int_a^b e^{-f(x)} \dd x \simeq \frac{e^{-f(b)}}{-f'(b)}
\end{equation}
instead, yielding equation (\ref{eq:tau0}) a few algebraic
manipulations later.

The calculation of the exact position of the superheated layer in the
limit where the disk is optically thick and the derivation of
equations (\ref{eq:thickzs}) and (\ref{eq:thickAs}) is slightly more
involved. To begin with, let's rewrite the gas density above the
photosphere in this region as
\begin{equation}
\rho(r,z) = \rho_{\rm e}(r)  e^{-\frac{z^2-z_{\rm e}^2}{2h_{\rm e}^2}}  \mbox{   ,   }
\end{equation}
where $\rho_{\rm e}(r)$ is the density at the photosphere, 
\begin{equation}
\rho_{\rm e}(r) = \rho_{\rm m}(r) \left(1-\frac{z_{\rm e}^2}{H^2} \right)^d \mbox{   .   }
\end{equation}

The equivalent calculation and manipulations that were used to obtain
equation (\ref{eq:tau0}) in the optically thin case can be used again
to yield this time
\begin{equation}
\tilde{Z} \rho_{\rm e} h_{\rm e} \kappa_{\rm V} e^{-\frac{z_{\rm s}^2-z_{\rm e}^2}{2h_{\rm e}^2}} = \frac{z^2_{\rm s}}{h_{\rm e}^2}  r\frac{\dd }{\dd r} \left(\frac{h_{\rm e}}{r} \right)  + \frac{h_{\rm e}}{r} \frac{\dd \ln (\tilde{Z}  \rho_{\rm e})}{\dd \ln r} + h_{\rm e} \frac{\dd}{\dd r} \left( \frac{z_{\rm e}^2}{2h_{\rm e}^2} \right)  \mbox{   .   }
\label{eq:tauthick}
\end{equation}
In the limit of a strongly opaque disk, $z_{\rm s} \gg h_{\rm e} $ and
$z_{\rm e} \gg h_{\rm e}$ (in other words, the atmosphere is much colder than the
mid-plane), so that we can simplify the above by neglecting the second
term in the RHS compared to the others. Now we use equation
(\ref{eq:he}) to simplify this to
\begin{equation}
\tilde{Z} \rho_{\rm e} h_{\rm e} \kappa_{\rm V} e^{-\frac{z_{\rm s}^2-z_{\rm e}^2}{2h_{\rm e}^2}} =  \frac{\theta^2}{\varepsilon_{\rm e}^{1/2}} \alpha  \mbox{   ,   }
\label{eq:A10}
\end{equation}
where $\alpha$ is defined by equation (\ref{eq:alpha}).

On the other hand, the total surface area of grains above the
superheated layer is given by
\begin{equation}
A_{\rm s} = \tilde{Z} \rho_{\rm e} \kappa_{\rm V} e^{-\frac{z_{\rm s}^2-z_{\rm e}^2}{2h_{\rm e}^2}}
\frac{h_{\rm e}^2}{z_{\rm s}} \mbox{   ,   }
\end{equation}
so that
\begin{equation}
A_{\rm s} = \frac{\theta^2}{\varepsilon_{\rm e}^{1/2}} \alpha \frac{h_{\rm e}}{z_{\rm s}} =
\frac{z_{\rm s}}{h} \alpha 
\end{equation}
on the assumption (which will be proved below) that $z_{\rm s} \sim H$.

To prove that $z_{\rm s} \sim H$ in the limit of a strongly opaque
disk we construct the quantity
\begin{equation}
\delta = \frac{z_{\rm s}^2 - z_{\rm e}^2}{H^2}  \mbox{   ,   }
\end{equation}
so that equation (\ref{eq:A10}) can be re-written as
\begin{equation}
\tilde{Z} \rho_{\rm m} \epsilon^{d} \kappa_{\rm V} e^{-\frac{\theta^2}{2\epsilon}
\delta} = \frac{\theta^2}{\varepsilon_{\rm e}} \frac{\alpha}{H}
\end{equation}
which can be solved for $\delta$ as
\begin{equation}
\delta = \frac{2\epsilon}{\theta^2} \ln\left(\frac{\epsilon^{n+1} \tilde{Z}
\kappa_{\rm V} \rho_{\rm m} H}{\alpha \theta^2}\right) =
\frac{2\epsilon}{\theta^2} \ln\left(\frac{\epsilon^{n+1} \tau_{\rm
V}^{\rm tot}}{I(d) \alpha \theta^2}\right) \mbox{   ,   }
\end{equation}
which shows that $\delta$ is proportional to
$\epsilon$, or in other words that $z_{\rm s}$ is close to $H$ when $z_{\rm e}$ is
close to $H$. 
\bigskip

\noindent {\it 3. Radial structure of the disk:} \\ When
solving for the radial structure of the disk, we are always
integrating ordinary differential equations of the kind
\begin{equation}
\frac{\dd }{\dd r} \left(\frac{h}{r} \right) = B \left(\frac{h}{r}
\right)^{a} r^b \mbox{   ,   }
\end{equation}
for which the general mathematical solution is
\begin{equation}
\left(\frac{h}{r} \right) = \left[ K + \frac{1-a}{b+1} B
r^{b+1}\right]^{\frac{1}{1-a}} \mbox{   .   }
\end{equation}
However, it is possible to set the integrating constant $K$ to zero on
physical grounds, since it would otherwise correspond to an unphysical
constant heating source in the disk. Once this is done, we simply get
\begin{equation}
h = \left[ \frac{1-a}{b+1} B\right]^{\frac{1}{1-a}} r^{\frac{b+1}{1-a} + 1}   \mbox{   .   } 
\end{equation}

\end{document}